\documentclass[12pt]{IEEEtran}
\normalsize

\usepackage{graphicx,amsmath,epsfig,times}
\usepackage{psfrag,cite}
\usepackage{amstext,amssymb,latexsym}
\usepackage{setspace}
\usepackage{algorithmic, algorithm}

\usepackage[latin5]{inputenc}

\allowdisplaybreaks

\onecolumn

\begin{document}
%
\title{A Generalized Framework on Beamformer Design and CSI Acquisition for Single-Carrier Massive MIMO Systems in Millimeter Wave Channels}
\author{Gokhan~M.~Guvensen,
~\IEEEmembership{Member,~IEEE},~and~
Ender~Ayanoglu,
~\IEEEmembership{Fellow,~IEEE}
\thanks{The authors are with the Center for Pervasive Communications and Computing (CPCC), Dept. of EECS, UC Irvine, CA, USA (e-mail: g.m.guvensen@uci.edu and ayanoglu@uci.edu.)}
}


\maketitle
\begin{abstract}
Recently, a two-stage beamforming concept under the name of Joint Spatial Division and Multiplexing (JSDM), a kind of divide-and-conquer approach based on statistical user-grouping, has been proposed to enable simplified system operations in massive MIMO. 
In this study, we establish a general framework on the reduced dimensional channel state information (CSI) estimation and pre-beamformer design for frequency-selective massive MIMO systems employing single-carrier (SC) modulation in time division duplex (TDD) mode by exploiting the joint angle-delay domain channel sparsity in millimeter (mm) wave frequencies (which is often characterized with limited scattering and hence correlatedness in the spatial domain). The main contribution of this work is threefold. First, by an inspiration from the user-grouping idea (in the JSDM framework), the reduced rank minimum mean square error (RR-MMSE) instantaneous CSI estimator, based on generic subspace projection taking the joint angle-delay power profile into account, is derived for spatially correlated wideband MIMO channels. Second, the statistical pre-beamformer design is considered for frequency-selective SC massive MIMO channels. We examine the dimension reduction and subspace (beamspace) construction on which the RR-MMSE estimation can be realized as accurately as possible. The generalized eigenvector beamspace (GEB) appears to be a nearly optimal pre-beamformer when the eigenspaces of different resolvable multi-path components are assumed to be nearly orthogonal. Finally, a spatio-temporal domain correlator type reduced rank channel estimator, as an approximation of the RR-MMSE estimate, is obtained by carrying out least square (LS) estimation in a proper reduced dimensional beamspace. It is observed that the proposed techniques show remarkable robustness to the pilot interference (or contamination) with a significant reduction in pilot overhead thanks to the subspace projection.
\end{abstract}


%

\section{Introduction}
\label{sec:introduction}
Massive multiple-input multiple-output (MIMO) systems, which are equipped with a large number of antenna elements at the base station (BS) to serve a relatively smaller number of user terminals (UTs) simultaneously, are believed to be one of the key technologies for next-generation cellular systems such as the upcoming 5G standard \cite{Larsson14,andrews14}.
With its potential large gains in spectral and energy efficiency, massive MIMO is especially promising for outdoor cellular systems operating at millimeter (mm) wave frequencies, where large antenna arrays can be packed into small form factors, and extremely large bandwidths are available for commercial use (e.g., up to 7 GHz in the 60 GHz band) \cite{ghosh14,ayanoglu14}. Thus, it is anticipated that massive MIMO systems in the mm wave range form an important part of 5G systems expected to support much larger, e.g., 1000 times faster data rates than the currently deployed standards \cite{andrews14}. 

Instantaneous channel state information (CSI) at BS is essential for massive MIMO transmission, since multi-user precoding at downlink or multi-user decoding at uplink necessitates accurate CSI in order to capitalize the aforementioned spatial diversity and multiplexing benefits of the channel \cite{swindlehurst14}. In practice, CSI is typically obtained with the assistance of the periodically inserted pilot signals \cite{swindlehurst14}. This brings the pilot overhead, namely, the amount of transmission resources (signaling dimensions per time-frequency channel coherence slot) consumed by the training data to be proportional to the number of active users in the system for uplink training, and the number of BS antennas for downlink training respectively \cite{ashikhmin11}. The acquisition of CSI in massive MIMO transmission has been studied extensively in the literature. One of the primary frameworks is the frequency division duplex (FDD) mode, where CSI is typically obtained through explicit downlink training and uplink (limited feedback) \cite{zoltowski14}. Since use of the FDD operation imposes a severe limit on the number of BS antennas due to the pilot overhead, alternatively, CSI at the BS can be acquired by means of uplink training in time division duplex (TDD) mode, where the uplink pilots provide the BS with downlink as well as uplink channel estimates simultaneously via leveraging the channel reciprocity \cite{swindlehurst14,ashikhmin11}.               

Although the TDD mode of operation eliminates the need for feedback and reduces the pilot overhead when compared to the FDD systems, the processing of the signals with very large dimensionality, the pilot interference, and the pilot overhead still constitute a bottleneck for the performance of massive MIMO transmission especially in mm wave frequencies even in TDD mode due to several reasons. First, in these systems, the elementary operations on the received signals such as matrix inversions and instantaneous CSI acquisition, multi-user precoding, decoding, equalization, adaptive spatial-temporal signal processing etc. become quickly infeasible with the increasing dimensions especially for large number of UTs. Second, for the conventional orthogonal training scheme in TDD mode \cite{swindlehurst14,ashikhmin11}, the pilot overhead would be prohibitively large for mm wave channels, where the signal-to-noise ratio (\textit{snr}) before beamforming is very small, and thus directional precoding/beamforming is inevitable to support longer outdoor links and to provide sufficient received signal power \cite{heath14,alkhateeb14}. However, the design of a directional beamformer is usually based on CSI. Moreover, utilizing orthogonal pilots among all users in the cell is one of the limiting factors on throughput in massive MIMO for high mobility scenarios where pilots must be transmitted more frequently, and for applications requiring low latency and short-packet duration \cite{you15,swindlehurst16}. These features are desirable for incorporating machine-type communications in next generation systems \cite{andrews14,buzzi14}. On the other hand, allowing pilot reuse (PR) among the \textit{intra-cell} UTs or non-orthogonal pilot assignment across the \textit{inter-cell} UTs (in neighboring cells) leads to the pilot interference \cite{you15} or pilot contamination \cite{ashikhmin11}, which undermine the value of MIMO systems in cellular networks.     
Therefore, in order to be able to exploit the advantages of massive MIMO communications, while overcoming the signal processing burden due to large dimensionality, pilot interference, and overhead bottleneck, some effective channel dimensionality reduction techniques, taking the slowly varying channel properties (\textit{long-term parameters}) (such as angles of arrival (AoAs), delays, and average power of the arriving waves) into account, must be employed. 

Recently, the \textit{two-stage beamforming} concept under the name of Joint Spatial Division and Multiplexing (JSDM) \cite{adhikary13,nam14} has been proposed to reduce the dimension of the MIMO channel effectively, and to enable massive MIMO gains and simplified system operations \cite{chen14,kim15}. Even though JSDM is suggested as an effective reduced-complexity two-stage downlink precoding scheme for multi-user MIMO systems in FDD mode initially, the idea of \textit{two-stage beamforming} (in \cite{adhikary13,nam14,kim15}) can be applied to both downlink and uplink transmission in TDD. JSDM can be seen as a divide-and-conquer approach considering the fact that the channel between a user and BS is spatially correlated. The key idea lies in \textit{user-grouping}, i.e., partitioning the user population supported by the serving BS into multiple groups each with approximately the same channel covariance eigenspaces. Then, one can decompose the MIMO beamformer at the BS into two steps via the use of spatial \textit{pre-beamformer}, which distinguishes \textit{intra-group} signals from other groups by suppressing the \textit{inter-group interference} while reducing the signaling dimension. The major complexity reduction in JSDM comes from the approach that the \textit{pre-beamformer} is properly designed based only on the \textit{long-term parameters} (described by using the second-order statistics of the channel) and not on the instantaneous CSI (which may vary on a much higher rate). In this case, the subsequent operations such as downlink multi-user precoding and uplink detection/decoding algorithms can be fulfilled based on the CSI of the \textit{effective} channel with significantly reduced dimensions thanks to the \textit{pre-beamforming} projection. At the same time, the training dimension necessary to learn the \textit{effective} channels of each UT is reduced considerably. Also, the JSDM scheme motivates the use of analog/digital MIMO architectures, specifically the so-called \textit{hybrid beamforming} \cite{ayach14,liu14,heath14,Noh16}, recently proposed as an alternative for fully digital precoding/decoding in mm wave, where efficient reconfigurable radio frequency (RF) architectures will be implemented at competitive cost, size, and energy in the near future. In the \textit{hybrid beamforming} architecture, the statistical \textit{pre-beamformer} (which depends on slowly varying parameters) may be implemented in the analog RF domain, while the multi-user MIMO precoding/decoding stage can be implemented by standard baseband processing.       

In this paper, we establish a general framework on the reduced dimensional CSI estimation and the \textit{pre-beamformer} design for frequency-selective massive MIMO systems employing single-carrier (SC) modulation in TDD mode by exploiting the \textit{channel sparsity} indicated by the joint \textit{angle-delay} domain power profile. The \textit{channel sparsity} \cite{bajwa09,swindlehurst16,chen16,caire16}, which becomes particularly relevant at mm wave frequencies, is observed in practical cellular systems, where the channels are often characterized with limited scattering and hence correlated in the spatial domain; the BS sees the incoming multi-path components (MPCs) under a very constrained angular range (AoA support), and the MPCs occur in clusters in the \textit{angle-delay} plane corresponding to the interaction with physical clusters of scatterers in the real world \cite{adhikary14}. Moreover, only MPCs, undergoing one or two reflections, can have significant power \cite{rappaport13,adhikary14}. On the other hand, the 5G systems, aimed to provide much higher throughput, will inevitably be broadband. Thus, the wideband massive MIMO channel is expected to be sparse both in angle and time (delay) domain. Recently, algorithms based on compressed sensing exploiting \textit{channel sparsity} gained attraction to realize channel estimation and reducing training overhead, e.g., \cite{rao14,shen16,heath15} and the references therein.
Nevertheless, the use of joint \textit{angle-delay} domain sparsity information is overlooked in the context of channel estimation with dimension reduction while taking the pilot interference and pilot overhead into account for SC systems in the TDD mode. 
Here, the reduced rank channel estimation problem based on generic subspace projection is handled by an inspiration from the JSDM framework, where the statistical \textit{pre-beamformer} is designed to reduce dimensionality and pilot overhead while mitigating \textit{inter-group interference} leading to pilot contamination (due to \textit{intra-} or \textit{inter-cell} UTs).            
The main contributions of this work are summarized as follows:
\begin{itemize}
\item
By an inspiration from the \textit{user-grouping} idea in the JSDM scheme, the reduced rank minimum mean square error (RR-MMSE) instantaneous CSI estimator, lying in the slowly varying second order statistics (given by the joint \textit{angle-delay} domain power profile), is derived for spatially correlated wideband MIMO channels. To the best of the authors' knowledge, the derivation of the RR-MMSE estimator, provided here, is
presented for the first time when the SC transmission with uplink training
in TDD mode is considered. 
\item
The statistical \textit{pre-beamformer} design is considered for frequency-selective SC massive MIMO channels. The fundamental approach here is to find a good subspace, on which the RR-MMSE channel estimation can be
realized as accurately as possible, so that a minimal performance
compromise in the subsequent statistical signal processing
operations after \textit{pre-beamforming} is provided. In this paper, we examine the dimension reduction problem by adopting several criteria based on the instantaneous CSI estimation accuracy. These criteria result in an equivalent optimization problem, and generalized eigenvector beamspace (GEB) appears to be a nearly optimal \textit{pre-beamformer} when the eigenspaces of different resolvable MPCs are assumed to be nearly orthogonal. Moreover, it is observed that RR-MMSE shows remarkable robustness to the pilot interference, and the significant reduction in pilot overhead is attained thanks to the dimension reducing subspace projection, which suppresses the \textit{inter-group} interfering signals.
\item
A spatio-temporal domain correlator type reduced rank channel estimator as a high \textit{snr} approximation of the RR-MMSE estimate is derived where the statistical (spatial) \textit{pre-beamforming} and (temporal) \textit{correlator} are applied in a successive manner. The key idea is to realize least square (LS) estimation in a proper reduced dimensional subspace so that the number of unknown parameters is reduced while capturing the intended part of the group signal and switching off the interference subspace, leading to pilot interference via \textit{pre-beamforming}. This approximate estimator is shown to be constructed based only on the \textit{pre-beamforming} matrix (determined by the support of the AoAs and delays of the MPCs) without necessitating the knowledge of the exact covariance matrices of the multi-path channel vector.  
\end{itemize} 

\section{System Model} 
\label{sec:sym_model}
We consider a cellular system based on massive MIMO transmission operating at mm wave bands in the TDD mode employing SC in which a BS, having $N$ antennas, serves $K$ single-antenna UTs. In order to reduce the overhead while acquiring the instantaneous CSI associated with massive MIMO, two-stage beamforming under the name of JSDM is adopted throughout this study. The main idea of JSDM scheme is based on partitioning the user population supported by the serving BS into multiple groups in order to enable massive MIMO gains and simplified system operations \cite{adhikary13,nam14}. As in JSDM-based transmission, $K$ users are partitioned into $G$ groups, where the $K_g$ users in group $g$ have statistically independent but identically distributed (\textit{i.i.d.}) channels \cite{adhikary13,nam14,kim15}%
\footnote{
Although JSDM is initially proposed as an effective reduced complexity two-stage downlink precoding scheme for multiuser massive MIMO systems in FDD mode \cite{adhikary13,nam14,chen14}, our focus here is on the reduced dimensional instantaneous channel acquisition technique and pre-beamformer design, with nearly optimal accuracy for uplink frequency-selective massive MIMO channels in TDD mode. This can be realized by exploiting the user-grouping idea inspired from the JSDM framework where a slowly-varying spatial correlation among the array elements exists.}.

At the beginning of every coherence interval, all users of the intended group $g$ 
transmit training sequences with length $T$. We assume a linear modulation (e.g., PSK or QAM) and a transmission over frequency-selective channel for all UTs with a slow evolution in time relative to the signaling interval (symbol duration). Under such conditions, the baseband equivalent received signal samples, taken at symbol rate ($W$) after pulse matched filtering, are expressed as%
\footnote{
Only the UTs, belonging to same group, are assumed to be synchronized for coherent uplink SC transmission. That is to say, only \textit{intra-group} synchronization is sufficient, and no synchronization and/or coordination is required between different group users (\textit{inter-group}). Note that assuming synchronization between uplink pilots of users can be regarded as a worst-case scenario from a \textit{intra-group} or \textit{inter-group} pilot interference point of view, since any lack of synchronization will tend to statistically decorrelate the pilots.}
\begin{equation}
\mathbf{y}_n=\underbrace{\sum_{\left\{k=1,\;g_k \in \Omega_g\right\}}^{K_g}\sum_{l=0}^{L_g-1}\mathbf{h}_l^{(g_k)}x_{n-l}^{(g_k)}}_
{\textrm{Intra-Group Signal}}\;
+ \underbrace{\sum_{\left\{\forall g'_k \in \Omega_{g'} 
\; \vert g' \neq g\right\}}
\left(\sum_{k=1}^{K_{g'}}\sum_{l=0}^{L_{g'}-1}
\mathbf{h}_l^{(g'_k)}x_{n-l}^{(g'_k)}\right)+\mathbf{n}_n}
_{\boldsymbol{\eta}_n^{(g)}: \textrm{Inter-Group Interference + AWGN}}
\label{eqn:multiuser_multipath_mimo_model}
\end{equation} 
\noindent for $n=0,\ldots,T-1$, where $\mathbf{h}_l^{(g_k)}$ is $N \times 1$ multi-path channel vector, namely, the array impulse response of the serving BS stemming from the $l^{th}$ multi-path component (MPC) of $k^{th}$ user in group $g$. It can be regarded as the discrete-time equivalent form of the channel response, and obtained after the symbol rate sampling of the impulse response, arising as the sum of the contributions from discrete MPCs, without any loss of information \cite{sykora00}. Here, $\left\{x_n^{(g_k)};\;-L_g+1 \leq n \leq T-1\right\}$ are the training symbols for the $k^{th}$ user in group $g$%
\footnote{
Training sequences are assumed to be non-orthogonal for synchronized \textit{intra-group} users for SC transmission in general. However, their temporal cross-correlation properties affect the accuracy of the CSI acquisition as will be apparent in the subsequent chapters. In addition to that, pilot reuse (PR) among \textit{inter-group} users is feasible thanks to the pre-beamforming yielding effective suppression for
\textit{inter-group} interference in spatial domain. This brings significant advantage in terms of pilot overhead which would be prohibitively large as the number of UTs become large, since utilizing orthogonal pilots among all users is one of the limiting factors on throughput in Massive MIMO \cite{you15} especially for applications requiring low latency and short-packet duration. These two features are desired for incorporating machine-type communications in next-generation cellular wireless systems \cite{andrews14}.},
$L_g$ is the channel memory of group $g$ multi-path channels%
\footnote{
In general, the multi-path channel is time unlimited, so that there are infinite number of nonzero $\frac{1}{W}$ spaced channel taps. However, it can be well 
approximated by finite number of nonzero channel coefficients \cite{sykora00}
as in (\ref{eqn:multiuser_multipath_mimo_model}).},
$\Omega_g$ is the set of all UTs belonging to group $g$ with cardinality $|\Omega_g|=K_g$, and $\left\{g_k\right\}_{k=1}^{K_g}$ are UT indices forming $\Omega_g$. The $L_g-1$ symbols at the start of the preamble, prior to the first observation at $n=0$, are the precursors. Training symbols are selected from a signal constellation $S \in \mathbb{C}$ and $\mathbb{E}\left\{|x_n^{(g_k)}|^2\right\}$ is set to $E_s$ for all $g_k$. 

In (\ref{eqn:multiuser_multipath_mimo_model}), $\mathbf{n}_n$ are the additive white Gaussian noise (AWGN) vectors during uplink pilot segment with spatially and temporarily 
\textit{i.i.d.} as $\mathcal{CN}\left(\mathbf{0},N_0\mathbf{I}_N\right)$, and $N_0$ is the noise power%
\footnote{
The received signal at BS is first pre-filtered by a brick-wall filter of proper bandwidth before unity gain pulse matched filtering and sampling at symbol rate ($W$) without information loss where the complex Gaussian baseband noise process is assumed to have circular symmetry and a flat power spectral density $N_0$ in the band of interest.}.
The first term of (\ref{eqn:multiuser_multipath_mimo_model})
is the transmitted signal of the intended group $g$, named as the \textit{intra-group} signal of group $g$ users. The 
second term, $\boldsymbol{\eta}_n^{(g)}$, namely the \textit{inter-group interference}, comprises of all the interfering signals, which stem from all inner or outer cell users belonging to different groups other than $g$. Finally, the average received signal-to-noise ratio (\textit{snr}) can be defined as 
$snr \triangleq \frac{E_s}{N_0}$%
\footnote{
It shows the maximum achievable \textit{snr} after beamforming when the beam is steered towards a point, i.e., angular location by assuming that the channel is normalized so that $\frac{1}{N}\frac{E_s}{N_0}$ can be seen as the average received \textit{snr} at each antenna element before beamforming.}.          


\subsection{Fundamental Assumptions on Signal and Channel Model}
\label{sec:assump_model}
In JSDM, where local scattering model is assumed, the BS sees the incoming MPCs under  a very constrained angular range, and the MPCs tend to occur in clusters on the \textit{angle-delay} plane, corresponding to the interaction with physical clusters of scatterers in the real world \cite{adhikary14,swindlehurst16,caire16}. Another important observation, which becomes particularly relevant at mm wave frequencies, is \textit{channel sparsity}. In other words, most of the channel power is concentrated in a finite region of angles or delays due to the limited scattering, and the number of significant MPCs is reduced to a much lower value than that for a microwave system operating in a similar environment \cite{bajwa09,adhikary14}. This sparsity can be resolved in the angle domain with the use of massive array architecture. As in JSDM-based systems, each resolvable MPC of the users, belonging to any group $g$, is assumed to span some particular angular sector in azimuth-elevation plane, capturing local scattering around the corresponding UT's angle of arrival (AoA). Then, their corresponding cross-covariance  matrices can be expressed in the form of
\begin{equation}
\mathbb{E}\left\{\mathbf{h}_l^{(g_k)}\left(\mathbf{h}_{l'}^{(g'_{k'})}\right)^H\right\}=\rho_l^{(g)}\mathbf{R}_l^{(g)}\delta_{gg'}\delta_{kk'}\delta_{ll'},
\textrm{ where } \sum_{l=0}^{L_g-1}\rho_l^{(g)}=1,\;
\operatorname{Tr}\left\{\mathbf{R}_l^{(g)}\right\}=1 
\label{eqn:mimo_multipath_correlations}
\end{equation}
\noindent by using the uncorrelated local scattering model where all MPCs are assumed to be mutually independent according to the well-known wide sense stationary uncorrelated scattering (WSSUS) model \cite{bello63,sykora00,swindlehurst16,adhikary14}, the multi-path channel vectors are uncorrelated with respect to $l$, and also mutually uncorrelated with that of the different users (independent of whether in the same group or not).  
In (\ref{eqn:mimo_multipath_correlations}), $\rho_l^{(g)}$ is the power delay profile (pdp) of the group $g$ multi-path channels, showing the average channel strength at each delay, and the auto-covariance of each MPCs in group $g$ is given by     
\begin{equation}
\mathbf{R}_l^{(g)}=\mathbf{U}_l^{(g)}\boldsymbol{\Lambda}_l^{(g)}
\left(\mathbf{U}_l^{(g)}\right)^H,\;l=0,\ldots,L_g-1, 
\label{eqn:mimo_MPC_covariance}
\end{equation}
\noindent where $\mathbf{U}_l^{(g)}$ is the $N \times r_{g,l}$ matrix of the eigenvectors corresponding to the $r_{g,l}$ non-zero or dominant eigenvalues of
$\mathbf{R}_l^{(g)}$, given as the diagonal elements of the diagonal $r_{g,l} \times r_{g,l}$ matrix $\boldsymbol{\Lambda}_l^{(g)}$ in (\ref{eqn:mimo_MPC_covariance}). 
In (\ref{eqn:mimo_multipath_correlations}), $\mathbf{R}_l^{(g)}$ can be considered as the common spatial covariance matrix of group $g$ UTs at $l^{th}$ delay. Under this model, $\mathbf{R}_l^{(g)}$ covers the predetermined sector with a particular center and angular spread (AS), where the diffuse radiation is included by considering intervals of angles for which the $l^{th}$ MPC have a continuum of non-resolvable components, each carrying infinitesimal scattered energy \cite{adhikary14}. That is to say, the $l^{th}$ MPC of group $g$ users stems from a particular scattering region for a given AoA support with respect to the BS.

In (\ref{eqn:mimo_MPC_covariance}), the effective rank of $\mathbf{R}_l^{(g)}$, namely, $r_{g,l}$ is expected to be much smaller than the number of array elements, $N$, since the \textit{channel sparsity} of the impulse response \cite{bajwa09} is pronounced at 
mm wave frequencies, where only MPCs undergoing one or two reflections can have significant power \cite{rappaport13,adhikary14}. The \textit{channel sparsity} is the source of significant correlation among the antenna array elements, which makes use of pre-beamforming very appealing in TDD or FDD modes in order to reduce the dimensionality of the multi-path channel%
\footnote{
Similar ideas would be applicable for downlink channel estimation in JSDM-based systems for the FDD mode (studied in \cite{zoltowski14,gao15}), provided UT is equipped with multiple antennas, in which case pre-beamformer would help simplify the instantaneous CSI acquisition and system operations at UTs, and reduce overheads significantly by suppressing \textit{inter-group interference} at the precoding stage.}.
This can be realized by exploiting the near-orthogonality of the eigenspaces of the MPCs of different user groups in joint \textit{angle-delay} domain.

When Rayleigh-correlated channel coefficients are assumed such
as $\mathbf{h}_l^{(g_k)}\thicksim 
\mathcal{CN}\left(\mathbf{0},\rho_l^{(g)}\mathbf{R}_l^{(g)}\right)$,
mutually independent across the users for all $g_k$, 
the \textit{Karhunen-Loeve} representation \cite{treesbook} of the multi-path channel vector belonging to the $k^{th}$ user in group $g$ is given as 
the following by using (\ref{eqn:mimo_MPC_covariance})
\begin{equation}
\mathbf{h}_l^{(g_k)}=\left(\rho_l^{(g)}\right)^{1/2}
\mathbf{U}_l^{(g)}\left(\boldsymbol{\Lambda}_l^{(g)}\right)^{1/2}
\mathbf{c}_l^{(g_k)},\;l=0,\ldots,L_g-1,
\label{eqn:KLT_expansion}
\end{equation}
\noindent where the entries of $\mathbf{c}_l^{(g_k)} \in \mathbb{C}^{r_{g,l} \times 1} \thicksim \mathcal{CN}\left(\mathbf{0},\mathbf{I}_{r_{g,l}}\right)$. 

We consider the following realistic assumptions related to the frequency-selective massive MIMO channel. The subsequent sections are based on these assumptions:
\begin{itemize}
\item
\textit{Block Fading} assumption, for which the channel is locally time-invariant over a packet duration, is adopted. Many existing cellular network standards based on pilot-aided channel estimation and coherent detection implicitly assume block fading \cite{goldsmithbook}.
\item
WSSUS model \cite{bello63,sykora00,adhikary14} is adopted for small-scale fading, namely, the normalized small-scale coefficients 
$\mathbf{c}_l^{(g_k)}$s in (\ref{eqn:KLT_expansion}) are assumed to be mutually independent, based on which (\ref{eqn:mimo_multipath_correlations}) is formed. 
\item
The cross-covariance matrix in (\ref{eqn:mimo_multipath_correlations})
is normalized so that the large-scale fading parameters such as path-loss and shadowing are incorporated into the average received signal-to-noise ratio (\textit{snr}). These parameters are assumed to be locally static, and the average channel strength can be easily learned over a long period of time. 
\item
The channel auto-covariance of each group in (\ref{eqn:mimo_MPC_covariance}) is slowly varying in time as the AoA of each user signal evolves depending on the user mobility, variation rate of the scattering environment characteristics, etc.\cite{you15,adhikary14,caire16,utschick05}. Also, this variation is known to be much slower than the actual Rayleigh fading process, then the WSSUS channel model is a local approximation with coherence time much larger than the small-scale fading coherence time. 
\item
The second order statistics of the multi-path channel vectors, namely, $\rho_l^{(g)}\mathbf{R}_l^{(g)}$ in (\ref{eqn:mimo_MPC_covariance}), varying at a much lower rate compared to the instantaneous CSI, can be estimated with guaranteed accuracy for all intended groups in practice, since there are enough time-frequency resources to be exploited for this purpose%
\footnote{
Algorithms for the covariance estimation or signal subspace tracking \cite{gershman03,marzetta11,chen16,caire16} could be utilized here to track the slow variations of the user channel covariance matrix together with the user grouping algorithms \cite{nam14,adhikary14} that partition users having approximately common subspace (characterizing group) in their MPCs. However, subspace tracking and user grouping algorithms are out of the scope of this work.}.       
\item
Mutual coupling, AS due to the diffuse scattering, AoA uncertainties of each UT stemming from the use of practical covariance estimation or tracking algorithms, user mobility, calibration errors or any other spatial correlation mismatches can be taken into account by choosing larger AoA support (dimension, i.e., $r_{g,l}$ in (\ref{eqn:KLT_expansion})) for each intended user group initially (to construct 
$\mathbf{R}_l^{(g)}$ in (\ref{eqn:mimo_MPC_covariance})), and then, 
$\mathbf{R}_l^{(g)}$s can be adaptively updated at a much lower rate compared to the instantaneous CSI learning.          
\end{itemize} 

In JSDM framework, users come in groups, either by nature or by the application of user grouping algorithms given in \cite{nam14}. In urban environments, it is typical to observe common clusters that create spatially correlated MPCs for many users. That is to say, when each user group is characterized by multiple scattering clusters, some of the clusters may significantly overlap in the \textit{angle-delay} plane. In this case, the user selection algorithms described in \cite{adhikary14} provide a set of user groups that can be served simultaneously in the same transmission resource
in (\ref{eqn:multiuser_multipath_mimo_model})%
\footnote{
As proposed under the JSDM framework, two user grouping techniques \cite{adhikary14} 
can be used to form (\ref{eqn:multiuser_multipath_mimo_model}):
1) \textit{Spatial Multiplexing}, that orthogonalizes two groups in the spatial domain via the pre-beamforming (i.e., suppressing common scatterers and/or inter-group interference), enables us to serve the two groups on the same transmission resource. 
2) \textit{Orthogonalization} serves the user groups, having common MPCs, in different channel transmission resources (time or frequency) by using 
pre-beamforming that allow all the channel eigenmodes (including the common scatterers) of each group to pass. 
First technique, yielding higher multiplexing gain, is shown to be effective at high-SNR regime, whereas the latter, providing full multi-path diversity gain, performs better at low-SNR regime \cite{adhikary14}.}.

Spatio-temporal covariance matrix of the \textit{inter-group interference} in (\ref{eqn:multiuser_multipath_mimo_model}) can be calculated by taking \textit{long-term} expectation over all MPCs $\mathbf{h}_{l'}^{(g'_{k'})}$s other than the ones belonging to group $(g)$ in the spatial domain, and transmitted symbols $x_{n'}^{(g'_{k'})}$s in the temporal domain. Considering the mutual independence across multi-path channel vectors (due to the uncorrelated scattering assumption for small-scale fading in WSSUS model) given by 
(\ref{eqn:mimo_multipath_correlations}), and considering that the transmitted symbols of different users are uncorrelated (including the data transmission period), i.e.,
$\mathbb{E}\left\{x_{n}^{(g_k)}\left(x_{n'}^{(g'_{k'})}\right)^H\right\}
=\gamma^{(g)}E_s\delta_{nn'}\delta_{gg'}\delta_{kk'}$, the following is obtained 
\begin{equation}
\mathbb{E}\left\{\boldsymbol{\eta}_n^{(g)}
\left(\boldsymbol{\eta}_{n'}^{(g)}\right)^H\right\}=
\mathbf{R}^{(g)}_{\boldsymbol{\eta}}\delta_{nn'},\textrm{ where }
\mathbf{R}^{(g)}_{\boldsymbol{\eta}} \triangleq 
E_s \left(\sum_{g' \neq g} \gamma^{(g')} K_{g'} \sum_{l=0}^{L_{g'}-1}
\rho_l^{(g')}\mathbf{R}_l^{(g')}\right)+N_0\mathbf{I}_N,
\label{eqn:InterGroupInterference} 
\end{equation}
\noindent and $\gamma^{(g')}$ for $g' \neq g$ can be regarded as the relative average received power at BS of \textit{inter-group} users normalized with that of the group $g$ users. In (\ref{eqn:InterGroupInterference}), 
$\gamma^{(g')}$s are accountable for the \textit{near-far} effect stemming from the fact that received signal strength of different UTs may differ significantly depending on their distance to the BS. Moreover, it is important to note that the $N \times N$ covariance matrix of the \textit{inter-group interference} $\mathbf{R}^{(g)}_{\boldsymbol{\eta}}$ in (\ref{eqn:InterGroupInterference}) consists of all the statistical information of the CSI in the spatial domain (i.e., AoA support) for all inner or outer cell users interfering with group $g$ users.
The interference covariance matrix, $\mathbf{R}^{(g)}_{\boldsymbol{\eta}}$ can be obtained by using the common spatial covariance of each intended group, namely, $\mathbf{R}_l^{(g)}$ in (\ref{eqn:mimo_MPC_covariance}) at each delay. 

The model in (\ref{eqn:multiuser_multipath_mimo_model}) can be applied to any SC-based MIMO setting such that both \textit{single-cell} or \textit{multi-cell}, where pilot contamination \cite{ashikhmin11} persists, can be considered%
\footnote{
Here, pilot contamination can be seen as the pilot interference stemming from inner or outer cell users belonging to groups apart from $g$ when they use training sequence non-orthogonal to that of the users in $g$.}.  
Regarding the \textit{multi-cell} scenario, if only statistical CSI coordination among cells is possible, covariance matrix of each MPC, belonging to user groups in neighboring cells, can be exchanged among the BSs. Then, the spatial covariance matrix of the interference  $\mathbf{R}^{(g)}_{\boldsymbol{\eta}}$ in (\ref{eqn:InterGroupInterference}) of each intended group can be calculated by taking the statistical CSI of the \textit{intra-cell} as well as the \textit{inter-cell} groups into account. Then, the statistical pre-beamforming stage can be realized to suppress all types of interfering sources accordingly%
\footnote{
Coordinated pilot allocation or scheduling algorithms for mitigating the \textit{intra-cell} or \textit{inter-cell} pilot contamination in \cite{you15,gesbert13} can be exploited, together with the selected user grouping technique and spatial pre-beamforming, in order to provide additional gain when there exist interfering groups having significant overlapping AoA support with that of the intended group $g$. This allows pilot reuse (PR) or non-orthogonal pilot sequences among \textit{intra-} or \textit{inter-cell} UTs apart from the 
\textit{intra-group} UTs of the intended groups, where the pilot length $T$ can be reduced substantially compared to the number of array elements in BS, $N$.}.     
Thus, the user grouping strategy with pre-beamforming inspired from JSDM \cite{nam14} can be seen as an appealing technique for SC uplink transmission in order to mitigate the pilot contamination effect considerably (in addition to significantly reduced system complexity). Moreover, in this setting, pilot reuse (PR) can be allowed among the \textit{inter-group} users, and thus the pilot overhead can be significantly reduced.      

Before concluding this section, it is better to emphasize one more time that AoAs and path strengths change only when the large scale geometry of the propagation between the transmitter and receiver significantly changes, thus their rate of change is significantly lower than that of the small-scale fading, namely, instantaneous CSI. In practice, AoAs and ASs (statistical CSI) of each UT can be determined by employing suitable compressed sensing tools \cite{alkhateeb14,heath15} where the sparse or low-rank nature of the MIMO channel at mm wave is taken into account%
\footnote{
Due to the sparse nature of the mm wave channel, compressed sensing (CS) algorithms \cite{alkhateeb14,heath15,caire16} can be employed to extract the statistical CSI, namely, the AoA support so that the covariance matrices of each MPC for all inner and outer cell groups, namely, $\mathbf{R}_l^{(g)}$'s in (\ref{eqn:mimo_MPC_covariance}) can be constructed. In general, these algorithms can be utilized as the initial acquisition tools of slowly-varying spatial correlation statistics necessary for instantaneous channel learning.}.
Therefore, long-term learning of AoA supports can be considered as the initial stage for a fine estimation of small-scale fading coefficients, namely, 
$\mathbf{c}_l^{(g_k)}$s in (\ref{eqn:KLT_expansion}) (instantaneous CSI), varying at a much higher rate than that of AoAs \cite{you15,caire16}.    


\subsection{Spatio-Temporal Domain Vector Definitions}
\label{sec:space_time_def}
In this paper, our main focus is on the uplink CSI acquisition that uses both \textit{angle-delay} domain sparsity information for spatially correlated MIMO channels described in Section \ref{sec:assump_model}. Before elaborating on the details of the estimation technique, we give the following matrix and vector definitions that will be useful in the subsequent chapters. First, the training matrix (or convolution matrix \cite{morelli00}), comprising of the transmitted pilots with the precursors for $k^{th}$ user in group $g$, is defined as%
\footnote{
If the BS has perfect knowledge of the sparsity pattern in \textit{angle-delay} domain such that some of the non-dominant MPCs are approximately zero, this a-priori information can be taken into account by simply setting $\rho_l^{(g)}$ in (\ref{eqn:mimo_multipath_correlations}) to zero for the corresponding delay with zero energy, or construct $\mathbf{X}_k^{(g)}$ in (\ref{eqn:training_matrix_user_k}) by extracting the columns corresponding to the multi-path channel taps possessing  
$\rho_l^{(g)}=0$.}       
\begin{equation}
\mathbf{X}_k^{(g)}\triangleq\left[\begin{array}{cccc}
x_0^{(g_k)} & x_{-1}^{(g_k)} & \cdots & x_{-L_g+1}^{(g_k)}\\
x_1^{(g_k)} & x_0^{(g_k)} & \cdots & x_{-L_g+2}^{(g_k)}\\
\vdots & \vdots & \ddots & \vdots \\
x_{T-1}^{(g_k)} & x_{T-2}^{(g_k)} & \cdots & x_{T-L_g}^{(g_k)}
\end{array}\right]_{T \times L_g}.
\label{eqn:training_matrix_user_k}
\end{equation}
\noindent The extended multi-path channel vector of the $k^{th}$ user, belonging to the intended group $g$, and its corresponding expansion coefficients after \textit{Karhunen-Loeve} Transform (KLT) in (\ref{eqn:KLT_expansion}) are given as
\begin{equation}
\mathbf{f}_k^{(g)}\triangleq\left[\begin{array}{c}
\mathbf{h}_0^{(g_k)} \\
\mathbf{h}_1^{(g_k)} \\
\vdots \\
\mathbf{h}_{L_g-1}^{(g_k)}\end{array}\right]_{NL_g \times 1},\;
\mathbf{b}_k^{(g)}\triangleq\left[\begin{array}{c}
\mathbf{c}_0^{(g_k)} \\
\mathbf{c}_1^{(g_k)} \\
\vdots \\
\mathbf{c}_{L_g-1}^{(g_k)}\end{array}\right]
_{\left(\sum_{l=0}^{L_g-1}r_{g,l}\right) \times 1}
\label{eqn:extended_ch_user_k}
\end{equation}
\noindent by concatenating all MPCs of the $k^{th}$ user in group $g$. Then, by using the vectors given in (\ref{eqn:multiuser_multipath_mimo_model}) and (\ref{eqn:extended_ch_user_k}), it will be useful to construct the following vectors that represents the whole received vector of signals at BS (in space-time domain) during training phase, the concatenated channel vector and its KLT coefficients (that include the channel parameters of all users in group $g$ to be estimated) respectively:   
\begin{align}
\mathbf{y}&\triangleq \operatorname{vec}\left\{\left[\begin{array}{cccc}
\mathbf{y}_0 & \mathbf{y}_1 & \cdots & \mathbf{y}_{T-1}
\end{array}\right]_{N \times T}\right\} \nonumber \\
\mathbf{h}^{(g)}&\triangleq \operatorname{vec}\left\{\left[\begin{array}{cccc}
\mathbf{f}_1^{(g)} & \mathbf{f}_2^{(g)} & \cdots & \mathbf{f}_{K_g}^{(g)} \end{array}\right]_{NL_g \times K_g}\right\} \label{eqn:vector_def2} \\
\mathbf{c}^{(g)}&\triangleq \operatorname{vec}\left\{\left[\begin{array}{cccc}
\mathbf{b}_1^{(g)} & \mathbf{b}_2^{(g)} & \cdots & \mathbf{b}_{K_g}^{(g)} \end{array}\right]
_{\left(\sum_{l=0}^{L_g-1}r_{g,l}\right) \times K_g}\right\}.
\nonumber
\end{align}
\noindent In a similar way, the \textit{inter-group interference} matrix with respect to group $g$ in space-time domain can be defined as 
\begin{equation}
\boldsymbol{\xi}^{(g)}\triangleq\operatorname{vec}
\left\{\left[\begin{array}{cccc}
\boldsymbol{\eta}_0^{(g)} & \boldsymbol{\eta}_1^{(g)} & \cdots & 
\boldsymbol{\eta}_{T-1}^{(g)} \end{array}\right]_{N \times T}\right\}.  \label{eqn:intergroup_space_time}
\end{equation}
\noindent Finally, the complete training matrix that consists of the training data of all users in group $g$ during the signaling interval $T$ is given by 
\begin{equation}
\mathbf{X}^{(g)}\triangleq \left[\begin{array}{cccc}
\mathbf{X}_1^{(g)} & \mathbf{X}_2^{(g)} & \cdots & \mathbf{X}_{K_g}^{(g)}
\end{array}\right]_{T \times K_gL_g}. 
\label{eqn:complete_training_matrix}
\end{equation}
The extended multi-path channel vector of group $g$ in 
(\ref{eqn:vector_def2}), carrying the complete CSI of all UTs in $g$, can be expressed in terms of KLT coefficients (small-scale fading) given in (\ref{eqn:KLT_expansion}) as  
\begin{align}
\mathbf{h}^{(g)}&= \underbrace{\left(\mathbf{I}_{K_g} \otimes \mathbf{V}\right)}
_{\triangleq \boldsymbol{\Upsilon}_{U}^{(g)}}\mathbf{c}^{(g)}
\textrm{ where }
\label{eqn:h_c_relation} \\
\mathbf{V} &\triangleq \operatorname{bdiag}\left[\left\{
\left(\rho_l^{(g)}\right)^{1/2}\mathbf{U}_l^{(g)}
\left(\boldsymbol{\Lambda}_l^{(g)}\right)^{1/2}\right\}
_{l=0}^{L_g-1}\right].
\label{eqn:V_def}
\end{align}
\noindent Here, (\ref{eqn:h_c_relation}) can be regarded as the generalized \textit{Karhunen-Loeve} expansion in the spatio-temporal domain (or \textit{angle-delay} domain) by using the corresponding eigenbasis given in (\ref{eqn:V_def}) with a-priori known channel power profile 
$\left\{\rho_l^{(g)}\boldsymbol{\Lambda}_l^{(g)}\right\}_{l=0}^{L_g-1}$ in the \textit{angle-delay} domain. Eigenbeams of group $g$, namely, $\left\{\mathbf{U}_l^{(g)}\right\}_{l=0}^{L_g-1}$ (with dimension $r_{g,l}$) do not need to be orthogonal to each other in general, i.e., overlapping can be observed between the MPCs at different delays. In (\ref{eqn:h_c_relation}), $\boldsymbol{\Upsilon}_{U}^{(g)} \triangleq \mathbf{I}_{K_g} \otimes \mathbf{V}$ is an $NK_gL_g \times 
K_g\left(\sum_{l=0}^{L_g-1}r_{g,l}\right)$ transform matrix (in the spatio-temporal domain) constructed by the eigenbasis of group $g$ at each delay. The KLT coefficients of group $g$ users in (\ref{eqn:h_c_relation}) are spatially and temporarily \textit{i.i.d.} with unity variance such that   
\begin{equation}
\mathbb{E}\left\{\mathbf{c}^{(g)}\left(\mathbf{c}^{(g)}\right)^H\right\}
=\mathbf{I}_{K_g\left(\sum_{l=0}^{L_g-1}r_{g,l}\right)}.
\label{eqn:Rc_eqn}
\end{equation}        

By using the matrices and vectors in spatio-temporal domain defined above, (\ref{eqn:multiuser_multipath_mimo_model}) can be expressed in a more compact matrix form
\begin{equation}
\mathbf{y} = \sum_{k=1}^{K_g}\left(\mathbf{X}_k^{(g)} \otimes 
\mathbf{I}_N\right)\mathbf{f}_k^{(g)}+\boldsymbol{\xi}^{(g)}
\label{eqn:y_eqn}
\end{equation}
\noindent where the covariance matrix of the spatio-temporal interference $\boldsymbol{\xi}^{(g)}$ with respect to $g$ can be obtained as  
\begin{equation}
\mathbf{R}^{(g)}_{\boldsymbol{\xi}} \triangleq 
\mathbb{E}\left\{\boldsymbol{\xi}^{(g)}
\left(\boldsymbol{\xi}^{(g)}\right)^H\right\}=
\mathbf{I}_T \otimes \mathbf{R}^{(g)}_{\boldsymbol{\eta}}
\label{eqn:space_time_noise}
\end{equation}
\noindent since $\boldsymbol{\eta}_n^{(g)}$s are uncorrelated in temporal domain due to (\ref{eqn:InterGroupInterference}).   

\subsection{Reduced Dimensional Spatio-Temporal Model for Sparse SC MIMO Channels}
\label{sec:reduced_dim_sys_model}
Next-generation wireless networks, composed of massive antenna arrays with several hundreds of receiving elements, utilize large dimensional received signal for uplink decoding or downlink precoding. In these systems, the elementary operations on the received signals such as matrix inversion and instantaneous CSI acquisition (with large pilot overheads) become quickly infeasible with the increasing dimensions especially for a large number of UTs. A good way to enable the processing of large-dimensional signals is the adoption of a pre-processing stage that captures the essence of the input at a reduced dimension. Inspired from the JSDM (or two-stage beamforming) framework \cite{adhikary13,kim15}, a spatial pre-beamformer, which is to be designed based only on statistical CSI, not on instantaneous CSI, is exploited to reduce the dimension of the signaling space. Thanks to the dimensionality reduction brought by the statistical pre-beamforming projection, instantaneous multi-path channel estimation (short-term) can be attained at considerably reduced complexity so that the multi-user precoding at downlink or multi-user decoding at uplink, necessitating instantaneous CSI for proper operation, can be fulfilled at reduced dimension with significantly reduced complexity.

In light of the discussion above, pre-beamforming is applied in order to distinguish \textit{intra-group} signal of group $g$ users from other groups by suppressing the \textit{inter-group interference}
while reducing the signaling dimension in (\ref{eqn:y_eqn}). At the pre-beamforming stage, a $DT$-dimensional space-time vector 
$\mathbf{y}^{(g)}$ is formed for all \textit{intra-cell} groups by a linear transformation through           
$\left(\mathbf{I}_T \otimes \left[\mathbf{S}_D^{(g)}\right]^H\right)$ matrix called $\left(\boldsymbol{\Upsilon}_S^{(g)}\right)^H$ as 
\begin{equation}
\mathbf{y}^{(g)} \triangleq \underbrace{\left(\mathbf{I}_T \otimes 
\left[\mathbf{S}_D^{(g)}\right]^H\right)}_
{\triangleq \left(\boldsymbol{\Upsilon}_S^{(g)}\right)^H}\mathbf{y},\;
g=1,\ldots,G,
\label{eqn:space_time_noise_reduced}
\end{equation}
\noindent where $\mathbf{S}_D^{(g)}$ is an $N \times D$ statistical pre-beamforming matrix that projects the $N$-dimensional received signal samples $\left\{\mathbf{y}_n\right\}_{n=0}^{T-1}$ in (\ref{eqn:multiuser_multipath_mimo_model}) on a suitable $D$-dimensional subspace in the spatial domain%
\footnote{
These motivate the use of analog/digital MIMO architectures recently proposed as an alternative for fully digital precoding/decoding in mm wave communication systems \cite{alkhateeb14,heath14}, since efficient reconfigurable radio frequency (RF) architectures will be implemented at competitive cost, size, and energy efficiency in near future \cite{adhikary14}. The advantage of implementing pre-beamforming in the analog RF domain is that the number of RF chains and analog-to-digital converters (ADCs) can be reduced so that the cost of baseband processing and baseband to RF modulation scales with the intermediate dimension which is $\sum_g \operatorname{rank}\left\{\mathbf{S}_D^{(g)}\right\}$, while the number of antennas $N$ can be very large.}. 
Our goal is to accomplish the dimension reduction with a small loss of instantaneous CSI of group $g$ UT estimation accuracy so that the following stages (after CSI acquisition) such as downlink precoding or uplink decoding can be realized with a close performance to that of the full dimensional case. The pre-beamformer design based only on the channel statistics for the aforementioned JSDM-based massive MIMO systems employing SC is considered in Section \ref{sec:beamformer_design}.

The output of the pre-beamformer $\mathbf{S}_D^{(g)}$ in (\ref{eqn:space_time_noise_reduced}) can be written explicitly as  
\begin{align}
\mathbf{y}^{(g)} &=
\left(\boldsymbol{\Upsilon}_S^{(g)}\right)^H\mathbf{y} \nonumber \\
&= \sum_{k=1}^{K_g}\left(\mathbf{X}_k^{(g)} \otimes 
\left[\mathbf{S}_D^{(g)}\right]^H\right)\mathbf{f}_k^{(g)}+
\underbrace{\left(\mathbf{I}_T \otimes \left[\mathbf{S}_D^{(g)}\right]^H\right)
\boldsymbol{\xi}^{(g)}}_{\boldsymbol{\xi}_D^{(g)}} \nonumber \\
&= \left(\mathbf{X}^{(g)} \otimes \left[\mathbf{S}_D^{(g)}\right]^H\right)\mathbf{h}^{(g)}+\boldsymbol{\xi}_D^{(g)} \nonumber \\
&= \left(\mathbf{X}^{(g)} \otimes \left[\mathbf{S}_D^{(g)}\right]^H\right)
\left(\mathbf{I}_{K_g} \otimes \mathbf{V}\right)\mathbf{c}^{(g)}+\boldsymbol{\xi}_D^{(g)} \nonumber \\
&= \boldsymbol{\Psi}_D^{(g)}\mathbf{c}^{(g)}+\boldsymbol{\xi}_D^{(g)}
\label{eqn:beamformer_out}
\end{align}
where
\begin{equation}
\boldsymbol{\Psi}_D^{(g)}=
\left(\mathbf{X}^{(g)} \otimes \left[\mathbf{S}_D^{(g)}\right]^H\right)
\left(\mathbf{I}_{K_g} \otimes \mathbf{V}\right).
\label{eqn:equivalent_channel}
\end{equation}
\noindent In (\ref{eqn:beamformer_out}), the second line follows from the Kronecker product rule $\left(\mathbf{A}_1 \otimes 
\mathbf{A}_2\right)\left(\mathbf{B}_1 \otimes \mathbf{B}_2\right)=
\left(\mathbf{A}_1\mathbf{B}_1\right) \otimes 
\left(\mathbf{A}_2\mathbf{B}_2\right)$ after substituting (\ref{eqn:y_eqn}) in its position, the third line follows from the definitions of the multi-path channel vector and the training matrix given in (\ref{eqn:vector_def2}) and (\ref{eqn:complete_training_matrix}) respectively, and the fourth line follows from the generalized KLT defined in spatio-temporal domain in (\ref{eqn:h_c_relation}). The expression in (\ref{eqn:beamformer_out}) is the equivalent spatio-temporal received signal representation of (\ref{eqn:multiuser_multipath_mimo_model}) after dimension reduction.
This expression, which contains all the relevant training information, channel statistics (sparsity information, AoA support, etc.) of \textit{intra-group} users and \textit{inter-group interference}, will be frequently used in the subsequent sections where the reduced dimensional channel estimator is constructed based on it.  

\section{Covariance-Based Reduced Rank Channel Estimation} 
\label{sec:LMMSE_est}
In this section, based on the model in (\ref{eqn:beamformer_out}),
the reduced dimensional linear minimum mean square error (LMMSE) channel estimator is derived while the side information lying in the second order statistics of the MPCs of each group is utilized.
As it is well-known, the LMMSE channel estimator is
often referred to as the \textit{Wiener filter}%
\footnote{
It is actually a Bayesian approach with a quadratic risk function, i.e., a conditional mean estimator \cite{treesbook} if \textit{intra-group} multi-path channel coefficients and \textit{inter-group} interference in (\ref{eqn:beamformer_out}) are jointly Gaussian distributed. In this case, the Bayesian estimator based on the maximum a posteriori (MAP) estimation rule, yielding the most probable value given the observation $\mathbf{y}^{(g)}$ after pre-beamforming, also coincides with the LMMSE estimate.}.
The reduced rank MMSE (RR-MMSE) estimate of the instantaneous CSI can be expressed in the following general form: 
\begin{equation}
\hat{\mathbf{h}}^{(g)}=
\boldsymbol{\Upsilon}_{U}^{(g)}
\hat{\mathbf{c}}^{(g)} 
=\boldsymbol{\Upsilon}_{U}^{(g)}\left(\mathbf{W}_{mmse,D}^{(g)}\right)^H
\mathbf{y}^{(g)}
=\underbrace{\boldsymbol{\Upsilon}_{U}^{(g)}\left(\mathbf{W}_{mmse,D}^{(g)}\right)^H
{\left(\boldsymbol{\Upsilon}_S^{(g)}\right)^H}}_
{\textrm{Reduced Rank Wiener Filter}}\mathbf{y}
\label{eqn:MMSE_1}
\end{equation}
\noindent where the RR-MMSE estimate of $\mathbf{h}^{(g)}$, namely,
$\hat{\mathbf{h}}^{(g)}$ is written in terms of the RR-MMSE estimate of $\mathbf{c}^{(g)}$, namely, $\hat{\mathbf{c}}^{(g)}$ (small-scale fading) by the KLT through $\boldsymbol{\Upsilon}_{U}^{(g)}$ matrix given in (\ref{eqn:h_c_relation}). This operation does not lead to any loss of information related to sufficient statistics, since the KLT matrix $\boldsymbol{\Upsilon}_{U}^{(g)}$ is one-to-one, i.e, full column rank. 

In (\ref{eqn:MMSE_1}), first, 
$\hat{\mathbf{c}}^{(g)}$ is formed by a reduced dimensional linear \textit{Wiener filter} (or MMSE) in the spatio-temporal domain through the
$\left(\mathbf{W}_{mmse,D}^{(g)}\right)^H$ matrix for 
group $g$ users after projecting (reducing the dimension) full dimensional observation $\mathbf{y}$ in (\ref{eqn:y_eqn}) onto a suitable subspace represented by 
$\left(\boldsymbol{\Upsilon}_S^{(g)}\right)^H$ (pre-beamforming) in (\ref{eqn:space_time_noise_reduced}). Then, the LMMSE estimate of the full-dimensional multi-path channel vector of group $g$, $\hat{\mathbf{h}}^{(g)}$, is constructed by transforming $\hat{\mathbf{c}}^{(g)}$ back to the original space by KLT through $\boldsymbol{\Upsilon}_{U}^{(g)}$. Thus, a general framework for the reduced dimensional channel estimation problem is established here (for general rank signal models) such that the overall reduced rank estimator given as the data processing chain: $\boldsymbol{\Upsilon}_{U}^{(g)}\left(\mathbf{W}_{mmse,D}^{(g)}\right)^H
\left(\boldsymbol{\Upsilon}_S^{(g)}\right)^H$, where the transform matrices $\left(\boldsymbol{\Upsilon}_S^{(g)}\right)^H$ (dimension-reducing subspace projection in the Kernel space) and 
$\boldsymbol{\Upsilon}_{U}^{(g)}$ (transforming back to the original space by KLT) are composed of different basis set in general. The \textit{Wiener filter} in (\ref{eqn:MMSE_1}) can be seen as the \textit{reduced rank approximation} of the full-dimensional \textit{Wiener filter}, i.e., 
$\left(\mathbf{W}_{full}^{(g)}\right)^H \approx 
\boldsymbol{\Upsilon}_{U}^{(g)}
\left(\mathbf{W}_{mmse,D}^{(g)}\right)^H
{\left(\boldsymbol{\Upsilon}_S^{(g)}\right)^H}$ where
the rank of the full-dimensional filter $\operatorname{rank}\left\{\mathbf{W}_{full}^{(g)}\right\}
=\operatorname{min}\left\{NK_gL_g,NT\right\}$ is reduced to $\operatorname{rank}\left\{\boldsymbol{\Upsilon}_{U}^{(g)}\left(\mathbf{W}_{mmse,D}^{(g)}\right)^H
{\left(\boldsymbol{\Upsilon}_S^{(g)}\right)^H}\right\}=
\operatorname{rank}\left\{\mathbf{W}_{mmse,D}^{(g)}\right\}
=\operatorname{min}\left\{K_g\sum_l r_{g,l},DT\right\}$. 
The transform matrices $\left(\boldsymbol{\Upsilon}_S^{(g)}\right)^H$ and $\boldsymbol{\Upsilon}_{U}^{(g)}$ do not depend on training data and instantaneous CSI, and are to be designed based on only long-term channel second order statistics, which brings significant complexity reduction especially when one considers the use of adaptive filtering and tracking algorithms. As it will be clear in the sequel, the proper design of pre-beamformer determining $\mathbf{S}_D^{(g)}$ is critical since there always exists some overlap among eigenspaces of different groups in the joint \textit{angle-delay} domain.               

\subsection{Joint Angle-Delay Domain Reduced Rank MMSE Estimator}
\label{sec:joint_angle_delay_mmse}
The reduced rank \textit{Wiener filter} $\left(\mathbf{W}_{mmse,D}^{(g)}\right)^H$ of group $g$, depending on the covariances of \textit{intra-group} signal and \textit{inter-group interference} (related to the joint \textit{angle-delay} domain sparsity information) in (\ref{eqn:MMSE_1}), can be obtained after the following mathematical steps by using the Kronecker product rule $\left(\mathbf{A}_1 \otimes 
\mathbf{A}_2\right)\left(\mathbf{B}_1 \otimes \mathbf{B}_2\right)=
\left(\mathbf{A}_1\mathbf{B}_1\right) \otimes 
\left(\mathbf{A}_2\mathbf{B}_2\right)$ successively such that        
\begin{align}
& \mathbf{W}_{mmse,D}^{(g)} =
\left(\mathbf{R}^{(g)}_{\mathbf{y}}\right)^{-1}\boldsymbol{\Psi}_D^{(g)} 
\nonumber \\
& =\left(\boldsymbol{\Psi}_D^{(g)}\left[\boldsymbol{\Psi}_D^{(g)}\right]^H+
\left(\boldsymbol{\Upsilon}_S^{(g)}\right)^H
\left(\mathbf{I}_T \otimes \mathbf{R}^{(g)}_{\boldsymbol{\eta}} \right)
\boldsymbol{\Upsilon}_S^{(g)}
\right)^{-1}\boldsymbol{\Psi}_D^{(g)}
\nonumber \\
& =\left\{\left(\mathbf{X}^{(g)} \otimes \left[\mathbf{S}_D^{(g)}\right]^H\right)\left(\mathbf{I}_{K_g} \otimes \mathbf{V}\mathbf{V}^H\right)
\left(\left[\mathbf{X}^{(g)}\right]^H \otimes \mathbf{S}_D^{(g)}\right)
\right. \nonumber \\
& \left. \qquad 
+\left(\mathbf{I}_T \otimes \left[\mathbf{S}_D^{(g)}\right]^H\right)
\left(\mathbf{I}_T \otimes \mathbf{R}^{(g)}_{\boldsymbol{\eta}} \right)
\left(\mathbf{I}_T \otimes \mathbf{S}_D^{(g)}\right)
\right\}^{-1}\boldsymbol{\Psi}_D^{(g)}
\nonumber \\
& =\left\{\sum_{l=0}^{L_g-1} \left(\mathbf{X}^{(g)} 
\left[\mathbf{I}_{K_g} \otimes \rho_l \mathbf{E}_{L_g,l}\right]\left[\mathbf{X}^{(g)}\right]^H \right) \otimes 
\left(\left[\mathbf{S}_D^{(g)}\right]^H \mathbf{R}_l^{(g)}
\mathbf{S}_D^{(g)}\right)+
\mathbf{I}_T \otimes \left(\left[\mathbf{S}_D^{(g)}\right]^H 
\mathbf{R}^{(g)}_{\boldsymbol{\eta}}\mathbf{S}_D^{(g)}\right)\right\}^{-1}
\boldsymbol{\Psi}_D^{(g)}.
\label{eqn:w_mmse_D}
\end{align}
\noindent In (\ref{eqn:w_mmse_D}), the first line follows from the solution of Wiener-Hopf equation \cite{treesbook} based on (\ref{eqn:beamformer_out}) defined in the spatio-temporal domain. Here, $\mathbf{R}^{(g)}_{\mathbf{y}}$ is defined as the covariance matrix of $\mathbf{y}^{(g)}$ in (\ref{eqn:beamformer_out}). Then, in the second line, the expression for $\mathbf{R}^{(g)}_{\mathbf{y}}$ is substituted into its place explicitly by using (\ref{eqn:Rc_eqn}), (\ref{eqn:space_time_noise}), and (\ref{eqn:beamformer_out}). The third line follows from (\ref{eqn:equivalent_channel}) by substituting $\boldsymbol{\Psi}_D^{(g)}$ and $\boldsymbol{\Upsilon}_S^{(g)}=\mathbf{I}_T \otimes \mathbf{S}_D^{(g)}$ into their places. Finally, the fourth line follows from the following useful expression obtained from (\ref{eqn:mimo_MPC_covariance}) and (\ref{eqn:V_def})     
\begin{equation}
\mathbf{V}\mathbf{V}^H =
\sum_{l=0}^{L_g-1} \rho_l \mathbf{E}_{L_g,l} \otimes 
\mathbf{R}_l^{(g)}
\label{eqn:Rv_eqn}
\end{equation}
\noindent where $\mathbf{E}_{L_g,l}$ is an $L_g \times L_g$ elementary diagonal matrix where all the entries except the $\left(l+1\right)^{th}$ diagonal one are zero. Then, substituting $\mathbf{W}_{mmse,D}^{(g)}$ in (\ref{eqn:w_mmse_D}) into the expression in (\ref{eqn:MMSE_1}) and using (\ref{eqn:equivalent_channel}), and after some straightforward steps noting that
$\left(\mathbf{A} \otimes \mathbf{B}\right)^{-1}=
\left(\mathbf{A}^{-1} \otimes \mathbf{B}^{-1}\right)$,
$\left(\mathbf{A} + \mathbf{B}\right)^{-1}=
\mathbf{B}^{-1}\left(\mathbf{I}+\mathbf{A}\mathbf{B}^{-1}\right)^{-1}$, and successive use of $\left(\mathbf{A}_1 \otimes 
\mathbf{A}_2\right)\left(\mathbf{B}_1 \otimes \mathbf{B}_2\right)=
\left(\mathbf{A}_1\mathbf{B}_1\right) \otimes 
\left(\mathbf{A}_2\mathbf{B}_2\right)$, the LMMSE estimate 
$\hat{\mathbf{h}}^{(g)}$ in (\ref{eqn:MMSE_1}) can be written 
explicitly as
\begin{align}
\hat{\mathbf{h}}^{(g)}&=
\left(\sum_{l=0}^{L_g-1} \left[\mathbf{I}_{K_g} \otimes 
\rho_l \mathbf{E}_{L_g,l} \right]\left[\mathbf{X}^{(g)}\right]^H
\otimes \mathbf{R}_l^{(g)} \mathbf{S}_D^{(g)} 
\left(\left[\mathbf{S}_D^{(g)}\right]^H 
\mathbf{R}^{(g)}_{\boldsymbol{\eta}}\mathbf{S}_D^{(g)}\right)^{-1} 
\right) \nonumber \\
& \qquad \left(\sum_{l=0}^{L_g-1}\mathbf{R}^{(g)}_{code}(l) \otimes
\left[\mathbf{SNR}^{(g)}_{mimo}(l)\right]^H+\mathbf{I}_{TD}\right)^{-1}
\mathbf{y}^{(g)}.
\label{eqn:h_mmse_est}
\end{align}
The matrices $\mathbf{SNR}^{(g)}_{mimo}(l)$ and $\mathbf{R}^{(g)}_{code}(l)$, 
appearing in (\ref{eqn:h_mmse_est}), are defined as 
\begin{align}
\mathbf{SNR}^{(g)}_{mimo}(l) &\triangleq 
\rho_l^{(g)}\left(\left[\mathbf{S}_D^{(g)}\right]^H 
\mathbf{R}^{(g)}_{\boldsymbol{\eta}}\mathbf{S}_D^{(g)}\right)^{-1}
\left(\left[\mathbf{S}_D^{(g)}\right]^H \mathbf{R}_l^{(g)}
\mathbf{S}_D^{(g)}\right), 
\label{eqn:snr_mimo_l} \\
\mathbf{R}^{(g)}_{code}(l) &\triangleq
\left(\mathbf{X}^{(g)} 
\left[\mathbf{I}_{K_g} \otimes \mathbf{E}_{L_g,l}\right]\left[\mathbf{X}^{(g)}\right]^H \right).
\label{eqn:r_code_l}
\end{align}

The matrices $\mathbf{SNR}^{(g)}_{mimo}(l)$ in (\ref{eqn:snr_mimo_l}) and 
$\mathbf{R}^{(g)}_{code}(l)$ in (\ref{eqn:r_code_l}) for $l=0,\ldots,L_g-1$ have useful properties explained in Appendix I.    
Briefly, the $D \times D$ positive semi-definite $\mathbf{SNR}^{(g)}_{mimo}(l)$ matrix in the spatial domain can be regarded as the generalized definition of the beamformer output \textit{snr} for general rank signal models \cite{gershman03,guvensen15}%
\footnote{
If the dimension of $\mathbf{S}_D^{(g)}$ is one, in this case $\mathbf{SNR}^{(g)}_{mimo}(l)$ is the \textit{snr} at the beamformer output when the beam is steered towards the AoA of $l^{th}$ MPC of group $g$ through pre-beamformer $\mathbf{S}_D^{(g)}$ for stochastic signals. The maximum value of \textit{snr} is attained when the \textit{Capon Beamformer} is utilized if the eigenspace of $l^{th}$ MPC is rank-1.\cite{Li03}}.
As it was shown in our previous work \cite{guvensen15}, $\operatorname{Tr}
{\left\{\mathbf{SNR}^{(g)}_{mimo}(l)\right\}}$ is actually the expected value of the point signal-to-interference noise ratio (\textit{sinr}) over the eigenspace (AoA support) of the $l^{th}$ MPC in group $g$, where the point \textit{sinr} is defined as the output \textit{snr} after beamforming when the beam is steered towards a point, i.e., angular location in the AoA support of $l^{th}$ MPC. In the temporal domain, the $T \times T$ positive semi-definite $\mathbf{R}^{(g)}_{code}(l)$ matrix is defined as the deterministic correlation matrix obtained from the columns of $\mathbf{X}^{(g)} 
\left[\mathbf{I}_{K_g} \otimes \mathbf{E}_{L_g,l}\right]$ where the columns with index $\left\{(l+1)+(k-1)L_g\right\}_{k=1}^{K_g}$ are the same as the $(l+1)^{th}$ column of the training matrix $\mathbf{X}_k^{(g)},\;k=1,\ldots,K_g$ in (\ref{eqn:training_matrix_user_k}), and the other elements are set to zero.
In (\ref{eqn:h_mmse_est}), the $\mathbf{SNR}^{(g)}_{mimo}(l)$ matrix is responsible for spatial processing only, utilizing the eigenspaces of the intended group $g$ at $l^{th}$ delay and the \textit{inter-group interference} after subspace projection onto $\mathbf{S}_D^{(g)}$. On the other hand, $\mathbf{R}^{(g)}_{code}(l)$ is responsible for temporal processing only, utilizing the temporal cross-correlation properties of the pilot sequences assigned to each UT.        

In order to harness the spatial multiplexing in each group, one consider the \textit{effective} multi-path channel vector of each group user $\mathbf{h}_{l,eff}^{(g_k)}$, seen after pre-beamforming as 
\begin{equation}
\mathbf{h}_{l,eff}^{(g_k)} \triangleq \left[\mathbf{S}_D^{(g)}\right]^H
\mathbf{h}_l^{(g_k)} \to
\mathbf{h}_{eff}^{(g)} \triangleq 
\left(\mathbf{I}_{K_gL_g} \otimes 
\left[\mathbf{S}_D^{(g)}\right]^H\right)
\mathbf{h}^{(g)}
\label{eqn:h_eff_def}
\end{equation}
\noindent from the definition of the extended multi-path channel for group $g$ in (\ref{eqn:extended_ch_user_k}) and (\ref{eqn:vector_def2}).
The subsequent stages at the transmitter or receiver, preceded by the pre-beamformer, fulfill \textit{intra-group} processing such as multi-user precoding (inner beamformer) at downlink or multi-user decoding at uplink in reduced dimensional subspace. These stages can access and utilize only this reduced dimensional \textit{effective} channel in (\ref{eqn:h_eff_def}). By using (\ref{eqn:h_mmse_est}) and the definition in (\ref{eqn:h_eff_def}), the RR-MMSE estimate of the \textit{effective} channel, seen after pre-beamforming, is constructed as     
\begin{align}
\hat{\mathbf{h}}_{eff}^{(g)} &\triangleq 
\left(\mathbf{I}_{K_gL_g} \otimes 
\left[\mathbf{S}_D^{(g)}\right]^H\right)\hat{\mathbf{h}}^{(g)}
\nonumber \\
&=
\left(\sum_{l=0}^{L_g-1} \left(\mathbf{X}^{(g)} \left[\mathbf{I}_{K_g} \otimes \mathbf{E}_{L_g,l}\right]\right) 
\otimes \mathbf{SNR}^{(g)}_{mimo}(l)
\right)^H \left(\sum_{l=0}^{L_g-1}\mathbf{R}^{(g)}_{code}(l) \otimes
\left[\mathbf{SNR}^{(g)}_{mimo}(l)\right]^H
+\mathbf{I}_{TD}\right)^{-1}
\mathbf{y}^{(g)}.
\label{eqn:h_eq_mmse_est}
\end{align}

In (\ref{eqn:h_eq_mmse_est}), it is observed that the complexity of calculating the instantaneous CSI estimate, stemming mainly from the matrix inversion, is substantially reduced thanks to the pre-beamformer 
which reduces the dimensionality with suitable projection subspace while increasing the \textit{snr} level (\textit{snr before beamforming}, i.e., $\frac{1}{N}\frac{E_s}{N_0}$ is typically very low especially at mm wave frequencies). The size of the matrix to be inverted in (\ref{eqn:h_mmse_est}) or (\ref{eqn:h_eq_mmse_est}) is independent of the number of array elements $N \gg D$. Moreover, the form of the RR-MMSE estimate in (\ref{eqn:h_eq_mmse_est}) is suitable to be used in decision-directed iterative channel estimation such that the decoded data can be exploited to construct $\mathbf{R}^{(g)}_{code}(l)$ in adaptive channel filtering or tracking mode.    

It appears that the form of multi-path channel estimate in joint \textit{angle-delay} domain in (\ref{eqn:h_eq_mmse_est}) is \textit{coupled} spatio-temporal processing in general, meaning that the spatial and temporal processing need to be accomplished jointly. As to the effectiveness of the proposed RR-MMSE estimator in terms of the pilot contamination effect, it is possible to attain considerable reduction in pilot interference (\textit{intra-} or \textit{inter-cell}) together with pilot overheads where the pilot length $T$ is kept small by allowing non-orthogonal sequences among \textit{intra-group} users and pilot reuse among \textit{inter-group} users. This can be achieved with the use of optimal joint spatio-temporal processing in (\ref{eqn:h_eq_mmse_est}), where the statistical pre-beamformer $\left[\mathbf{S}_D^{(g)}\right]^H$ suppresses the \textit{inter-group} interfering signals leading to pilot interference, and $\left(\mathbf{X}^{(g)} \left[\mathbf{I}_{K_g} \otimes \mathbf{E}_{L_g,l}\right]\right)^H$ is a kind of temporal (Rake-type) correlator used to differentiate different MPCs having overlapping AoA support in the spatial domain.           
In (\ref{eqn:h_eq_mmse_est}), $\left(\mathbf{X}^{(g)} \left[\mathbf{I}_{K_g} \otimes \mathbf{E}_{L_g,l}\right]\right)^H$ simply selects the $l^{th}$ delayed signal, i.e., places a temporal finger on the $l^{th}$ temporal diversity path for all $K_g$ \textit{intra-group} users, while $\mathbf{SNR}^{(g)}_{mimo}(l)$ is accountable for applying optimum spatial weights given the power profile $\left\{\rho_l^{(g)}\boldsymbol{\Lambda}_l^{(g)}\right\}_{l=0}^{L_g-1}$ in the \textit{angle-delay} domain after (spatial) beamforming $\left[\mathbf{S}_D^{(g)}\right]^H$ in order to suppress the \textit{inter-group interference} effectively.     


The derivation of the RR-MMSE estimator provided here is presented for the first time when the SC uplink transmission in TDD mode is considered for frequency-selective 
multi-user spatially correlated MIMO channels with a given long-term joint \textit{angle-delay} power profile. Different than the previous low-rank LMMSE approaches in \cite{utschick05,chen10,gandara11}, the RR-MMSE estimator here can be interpreted as the reduced rank approximation of the optimal spatio-temporal Wiener filter in (reduced dimensional) transformed domain by using two different generic transform basis sets for projection onto a suitable subspace (pre-beamformer) and KLT while there exists overlap between eigenspaces of different groups in joint \textit{angle-delay} domain in general. Thus, for the model here, 
which provides a general description for massive MIMO based transmission employing SC in frequency-selective fading, the proposed covariance-based reduced rank estimator here, confirm, compare, and complement many previous works, where the pilot interference due to the use of non-orthogonal pilots in \textit{intra-} or \textit{inter-cell} users persists, by changing several system and model parameters.        

\subsection{Angle Domain Reduced Rank MMSE Estimator}
\label{sec:angle_only_mmse}
One can consider the following approximation of (\ref{eqn:h_eq_mmse_est}) by 
assuming that the MPCs of each group have the same AoA support (common angular sector) with the following covariance matrix: 
\begin{equation}
\mathbf{R}^{(g)}_{sum} \triangleq \sum_{l=0}^{L_g-1} \rho_l^{(g)}\mathbf{R}_l^{(g)}.
\label{eqn:Rsum_g}
\end{equation}
\noindent This corresponds to the use of \textit{angular} information only when all the AoA supports of each MPC, belonging to the same group, are unified as in (\ref{eqn:Rsum_g}). In (\ref{eqn:snr_mimo_l}), by replacing $\rho_l^{(g)}\mathbf{R}_l^{(g)}$ with $\mathbf{R}^{(g)}_{sum}$, one can get the following approximation for the effective channel estimate in (\ref{eqn:h_eq_mmse_est}): 
\begin{equation}
\hat{\mathbf{h}}_{eff,2}^{(g)}=
\left(\mathbf{X}^{(g)} \otimes \mathbf{SNR}^{total,(g)}_{mimo}\right)^H 
\left(\mathbf{R}^{(g)}_{code} \otimes
\mathbf{SNR}^{total,(g)}_{mimo}+\mathbf{I}_{TD}\right)^{-1}
\mathbf{y}^{(g)}
\label{eqn:h_eq_mmse_approx}
\end{equation}
\noindent where $\mathbf{SNR}^{total,(g)}_{mimo}$ and $\mathbf{R}^{(g)}_{code}$ matrices are defined accordingly as  
\begin{align}
\mathbf{SNR}^{total,(g)}_{mimo} &\triangleq  \left(\left[\mathbf{S}_D^{(g)}\right]^H 
\mathbf{R}^{(g)}_{\boldsymbol{\eta}}\mathbf{S}_D^{(g)}\right)^{-1}
\left(\left[\mathbf{S}_D^{(g)}\right]^H 
\mathbf{R}^{(g)}_{sum}
\mathbf{S}_D^{(g)}\right),
\label{eqn:snr_mimo_total} \\
\mathbf{R}^{(g)}_{code} &\triangleq
\mathbf{X}^{(g)} \left[\mathbf{X}^{(g)}\right]^H.
\label{eqn:r_code_total}
\end{align}
\noindent This estimator is called the \textit{angle} domain RR-MMSE estimator, which will be used in pre-beamformer design and performance comparison in the sequel.

\section{Nearly Optimal Beamformer Design}
\label{sec:beamformer_design}
In this section, we consider the pre-beamformer $\mathbf{S}_D^{(g)}$ design based only on the second order channel statistical information of user groups in (\ref{eqn:multiuser_multipath_mimo_model}). The problem of statistical pre-beamformer design is handled for two-stage beamforming framework using JSDM in several recent studies \cite{adhikary13,chen14,liu14,kim15} where the block diagonalization (BD) algorithms were investigated in order to reduce the dimensionality for simplified system operation in multi-user precoding at downlink or enabling massive MIMO gains in FDD mode. 
In \cite{adhikary13}, the BD is obtained by projecting the dominant eigenvectors of the desired group channel covariance matrix on to the null space of the dominant eigenspace of all other groups. In \cite{chen14}, the pre-beamformer is constructed from the minimization of the \textit{inter-group interference} power minus the weighted \textit{intra-group} signal power. In \cite{liu14}, the pre-beamformer is considered as a part of the phase-only analog precoding stage which was obtained as a set of columns chosen from a discrete Fourier transform (DFT) matrix. In \cite{kim15}, the average signal-to-leakage plus noise ratio (\textit{slnr}), which is the ratio of \textit{intra-group} signal power received at the intended UT to the \textit{intra-group} signal received by the undesired \textit{inter-group} UTs for downlink transmission, is adopted as an optimization criterion. Then, the statistical pre-beamformer is obtained as a result of the trace quotient problem (TQP). All of these recent studies, related to the pre-beamformer design, consider flat-fading spatially-correlated massive MIMO channel, whereas in this paper, the pre-beamformer design is tackled for frequency-selective massive MIMO systems employing SC in TDD mode where the joint \textit{angle-delay} domain power profile of the channel is taken into account in general. 

Our goal is to find a good subspace (spanned by the columns of $\left(\mathbf{S}_D^{(g)}\right)$ matrix) on which the reduced dimensional instantaneous channel estimation can be realized as accurately as possible, so that a minimal performance compromise in the subsequent statistical signal processing operations after beamforming is provided. This approach, adopting CSI estimation accuracy after pre-beamforming as a performance measure with the use of more general joint \textit{angle-delay} channel profile, is completely different than the previous works in the massive MIMO literature.               

\subsection{Beamformer Design Criteria}
\label{sec:design_criteria}
It is well-known that the minimal sufficient statistics of the detection and estimation theory reduce the dimension of the input at no loss of information \cite{treesbook}. Unfortunately, it is difficult to find non-trivial sufficient statistics, yet the minimal one, in many problems and the derivation of useful statistics remains as an important challenge for such problems. In this paper, we examine the dimension reduction problem from three different viewpoints based on the instantaneous CSI estimation accuracy. These criteria result in an equivalent optimization problem yielding the optimal dimension-reducing subspace.  

\subsubsection{Reconstruction Error Minimizing Subspace}
\label{sec:error_minimizing_subspace:}
The reduced rank Wiener filtering in (\ref{eqn:MMSE_1}) can be seen as the data reconstruction process from noisy observations after dimension reduction. A legitimate goal is the minimization of the reconstruction error according to a criterion. If we denote the reconstruction error vector with 
$\mathbf{e}^{(g)} \triangleq \mathbf{h}^{(g)}-\hat{\mathbf{h}}^{(g)}$ for group $g$ channels, the covariance matrix of error $\mathbf{R_e}^{mmse}$ can be calculated as follows:    
\begin{align}
& \mathbf{R_e}^{mmse} \triangleq \mathbb{E}\left\{\left(\mathbf{h}^{(g)}-
\hat{\mathbf{h}}^{(g)}\right)
\left(\mathbf{h}^{(g)}-\hat{\mathbf{h}}^{(g)}\right)^H\right\}
\nonumber \\
& =\boldsymbol{\Upsilon}_{U}^{(g)} \mathbb{E}\left\{\left(\mathbf{c}^{(g)}-\hat{\mathbf{c}}^{(g)}\right)
\left(\mathbf{c}^{(g)}-\hat{\mathbf{c}}^{(g)}\right)^H\right\}
\left(\boldsymbol{\Upsilon}_{U}^{(g)}\right)^H
\nonumber \\
& =\boldsymbol{\Upsilon}_{U}^{(g)}\left[\mathbf{I}_
{K_g\left(\sum_{l=0}^{L_g-1}r_{g,l}\right)}-
\left(\boldsymbol{\Psi}_D^{(g)}\right)^H\mathbf{W}_{mmse,D}^{(g)}\right]
\left(\boldsymbol{\Upsilon}_{U}^{(g)}\right)^H
\nonumber \\
& =\sum_{l=0}^{L_g-1} \left[\mathbf{I}_{K_g} \otimes \rho_l \mathbf{E}_{L_g,l}\right] \otimes \mathbf{R}_l^{(g)} 
-\boldsymbol{\Upsilon}_{U}^{(g)}
\left(\boldsymbol{\Psi}_D^{(g)}\right)^H
\left(\mathbf{R}^{(g)}_{\mathbf{y}}\right)^{-1}\boldsymbol{\Psi}_D^{(g)}
\left(\boldsymbol{\Upsilon}_{U}^{(g)}\right)^H
\nonumber \\
& =\mathbf{R}_{full}^{(g)}-\mathbf{R}_{full}^{(g)}\mathbf{F}_s^{(g)}
\left(\mathbf{R}^{(g)}_{\mathbf{y}}\right)^{-1}
\left(\mathbf{F}_s^{(g)}\right)^H\mathbf{R}_{full}^{(g)}
\label{eqn:mmse_error_cov}
\end{align}
where
\begin{align}
\mathbf{R}^{(g)}_{\mathbf{y}} &=
\left(\mathbf{F}_s^{(g)}\right)^H\mathbf{R}_{full}^{(g)}\mathbf{F}_s^{(g)}+
\left(\boldsymbol{\Upsilon}_S^{(g)}\right)^H
\mathbf{R}^{(g)}_{\boldsymbol{\xi}}
\boldsymbol{\Upsilon}_S^{(g)}, \nonumber \\
\mathbf{R}_{full}^{(g)} &\triangleq 
\mathbb{E}\left\{\mathbf{h}^{(g)}\left(\mathbf{h}^{(g)}\right)^H\right\}=
\sum_{l=0}^{L_g-1} 
\left[\mathbf{I}_{K_g} \otimes \mathbf{E}_{L_g,l}\right] \otimes \rho_l \mathbf{R}_l^{(g)},
\nonumber \\
\mathbf{F}_s^{(g)} &\triangleq \left(\left[\mathbf{X}^{(g)}\right]^H 
\otimes \mathbf{S}^{(g)}_D\right), \nonumber \\
\boldsymbol{\Upsilon}_S^{(g)} &\triangleq
\left(\mathbf{I}_T \otimes \mathbf{S}^{(g)}_D\right).
\label{eqn:matrix_def_error_cov}
\end{align}
\noindent In (\ref{eqn:mmse_error_cov}), the second line follows from the KLT expression in (\ref{eqn:h_c_relation}) and (\ref{eqn:MMSE_1}). The third line follows from the estimation error covariance calculation for $\mathbf{c}^{(g)}$ when Wiener filter $\mathbf{W}_{mmse,D}^{(g)}$ in 
(\ref{eqn:MMSE_1}) is applied to $\mathbf{y}^{(g)}$ given by  (\ref{eqn:beamformer_out}), and considering the fact that error vector is uncorrelated with the observation $\mathbf{y}^{(g)}$ in $D$-dimensional subspace. In the fourth line, the first term, named $\mathbf{R}_{full}^{(g)}$, is calculated by using the spatio-temporal KLT definition in (\ref{eqn:h_c_relation}) and the expression in (\ref{eqn:Rv_eqn}), and the second term is obtained by substituting $\mathbf{W}_{mmse,D}^{(g)}$ into its place. Finally, the fifth line follows from using (\ref{eqn:equivalent_channel}) and (\ref{eqn:Rv_eqn}) after the successive use of the Kronecker product rule $\left(\mathbf{A}_1 \otimes 
\mathbf{A}_2\right)\left(\mathbf{B}_1 \otimes \mathbf{B}_2\right)=
\left(\mathbf{A}_1\mathbf{B}_1\right) \otimes 
\left(\mathbf{A}_2\mathbf{B}_2\right)$. Moreover, it is important to note that the inverse of $\mathbf{R_e}^{mmse}$ in (\ref{eqn:mmse_error_cov}) is actually the Fisher information matrix \cite{treesbook}.              

\underline{\textit{Error Volume}}: The minimization of the estimation error volume, namely, the determinant of $\mathbf{R_e}^{mmse}$ in (\ref{eqn:mmse_error_cov}), can be regarded as one of the important objectives on which $D$-dimensional subspace 
$\mathbf{S}^{(g)}_D$ is optimized. If one takes the determinant of both parts in (\ref{eqn:mmse_error_cov}), the following expression is obtained  
\begin{equation}
\operatorname{det}\left(\mathbf{R_e}^{mmse}\right)=
\frac{\operatorname{det}\left(\mathbf{R}_{full}^{(g)}\right)}
{\operatorname{det}\left(\mathbf{I}_{TD}+\sum_{l=0}^{L_g-1}\mathbf{R}^{(g)}_{code}(l) \otimes \mathbf{SNR}^{(g)}_{mimo}(l)\right)} 
\label{eqn:error_volume}
\end{equation}
\noindent where $\mathbf{R}_{full}^{(g)}$ defined in (\ref{eqn:matrix_def_error_cov}) can be seen as a priori error volume of $\mathbf{h}^{(g)}$ before the training period. The mathematical details of this derivation can be found in Appendix II.  
  
\underline{\textit{Normalized Mean Square Error}}: The normalized mean square error (\textit{nMSE}) covariance can be defined as the estimation error covariance matrix of the KLT coefficients $\mathbf{c}^{(g)}$ in (\ref{eqn:h_c_relation}) 
\begin{equation}
\mathbf{nMSE}^{(g)} \triangleq \mathbb{E}\left\{\left(\mathbf{c}^{(g)}
-\hat{\mathbf{c}}^{(g)}\right)
\left(\mathbf{c}^{(g)}-\hat{\mathbf{c}}^{(g)}\right)^H\right\}.
\label{eqn:nmse_error_cov_def}
\end{equation}
\noindent Then, the trace of $\mathbf{nMSE}^{(g)}$, as an alternative objective function, can be obtained in the following compact form as 
\begin{equation}
\operatorname{Tr}\left\{\mathbf{nMSE}^{(g)}\right\} \triangleq
\operatorname{Tr}\left\{
\left(\sum_{l=0}^{L_g-1}\mathbf{R}^{(g)}_{code}(l) \otimes
\mathbf{SNR}^{(g)}_{mimo}(l)+\mathbf{I}_{TD}\right)^{-1}\right\}
+\left(K_g\left(\sum_{l=0}^{L_g-1}r_{g,l}\right)-TD\right)
\label{eqn:nmse_error_cov}
\end{equation}
\noindent after some mathematical manipulations given in Appendix II. We would like to note that the scalar version of the relation (\ref{eqn:nmse_error_cov}) for $T=1,\;D=1$, and $K_g=1,\;L_g=1$ that is $nMSE=1/(1+snr)$, utilized in the analysis
of communication systems \cite{palomar03,schubert08}.  

\subsubsection{Mutual Information Preserving Subspace}
\label{sec:mutual_info}
The mutual information between observation $\mathbf{h}^{(g)}$ in 
(\ref{eqn:vector_def2}) and $\mathbf{y}$ in (\ref{eqn:y_eqn})  
can be written as $I\left(\mathbf{h}^{(g)};\mathbf{y}\right)$ \cite{coverbook}. From the data processing inequality, it is known that 
$I\left(\mathbf{h}^{(g)};\mathbf{y}\right) \geq
I\left(\mathbf{h}^{(g)};\mathbf{y}^{(g)}\right)$. 
The equality is only satisfied if $\mathbf{y}^{(g)}$ in (\ref{eqn:space_time_noise_reduced}) is the sufficient statistic with respect to the joint probability distribution of $\mathbf{h}^{(g)}$ and $\mathbf{y}$. Here, our goal is to find an $N \times D$ $\mathbf{S}^{(g)}_D$ pre-beamformer matrix so that $I\left(\mathbf{h}^{(g)};\mathbf{y}^{(g)}\right)$ is as close as possible to $I\left(\mathbf{h}^{(g)};\mathbf{y}\right)$ under the given dimension reduction constraint. The problem can also be stated as the preservation of the mutual information with a linear transformation under a rank constraint.
By assuming that both $\mathbf{h}^{(g)}$ and $\mathbf{y}^{(g)}$ are jointly Gaussian in (\ref{eqn:beamformer_out}), their joint density can be easily expressed in terms of the covariance and cross-covariance of these two vectors \cite{telatar99}. The mutual information between 
$\mathbf{h}^{(g)}$ and $\mathbf{y}^{(g)}$ in (\ref{eqn:beamformer_out}) can be compactly obtained by using (\ref{eqn:Rc_eqn}) and the covariance matrix of \textit{inter-group interference} $\mathbf{R}^{(g)}_{\boldsymbol{\xi}}$ in (\ref{eqn:space_time_noise}) as follows 
\begin{align}
& I\left(\mathbf{h}^{(g)};\mathbf{y}^{(g)}\right)=
I\left(\mathbf{c}^{(g)};\mathbf{y}^{(g)}\right) \nonumber \\
& =\operatorname{log}\left\{\operatorname{det}\left( \mathbf{I}_{TD}+
\left[\left(\boldsymbol{\Upsilon}_S^{(g)}\right)^H
\mathbf{R}^{(g)}_{\boldsymbol{\xi}}
\boldsymbol{\Upsilon}_S^{(g)}\right]^{-1} 
\boldsymbol{\Psi}_D^{(g)}\left[\boldsymbol{\Psi}_D^{(g)}\right]^H
\right)\right\} \nonumber \\
& =\operatorname{log}\left\{\operatorname{det}\left( \mathbf{I}_{TD}+
\left[\mathbf{I}_{T} \otimes 
\left(\left[\mathbf{S}_D^{(g)}\right]^H 
\mathbf{R}^{(g)}_{\boldsymbol{\eta}}\mathbf{S}_D^{(g)}\right)^{-1}\right]
\right. \right. \nonumber \\
& \left. \left. \qquad \qquad 
\sum_{l=0}^{L_g-1} \left(\mathbf{X}^{(g)} 
\left[\mathbf{I}_{K_g} \otimes \rho_l \mathbf{E}_{L_g,l}\right]\left[\mathbf{X}^{(g)}\right]^H \right) 
\otimes 
\left(\left[\mathbf{S}_D^{(g)}\right]^H \mathbf{R}_l^{(g)}
\mathbf{S}_D^{(g)}\right) \right)\right\} \nonumber \\
& =\operatorname{log}\left\{\operatorname{det}\left( \mathbf{I}_{TD}+\sum_{l=0}^{L_g-1}\mathbf{R}^{(g)}_{code}(l) \otimes 
\mathbf{SNR}^{(g)}_{mimo}(l) \right) \right\}.
\label{eqn:mutual_inf_S_D}
\end{align}
\noindent In (\ref{eqn:mutual_inf_S_D}), the first line follows from the fact that the KLT matrix in (\ref{eqn:h_c_relation}) is full column rank, and the second line is written based on the reduced dimensional system model in (\ref{eqn:beamformer_out}) after following a similar way to the calculation of channel capacity for a non-fading MIMO channel \cite{stoica05}. The third line follows from the successive use of the Kronecker product rule $\left(\mathbf{A}_1 \otimes 
\mathbf{A}_2\right)\left(\mathbf{B}_1 \otimes \mathbf{B}_2\right)=
\left(\mathbf{A}_1\mathbf{B}_1\right) \otimes 
\left(\mathbf{A}_2\mathbf{B}_2\right)$ after substituting
$\boldsymbol{\Psi}_D^{(g)}$ in (\ref{eqn:equivalent_channel}) in its place and using (\ref{eqn:Rv_eqn}). Finally, by using the Kronecker product rule, the fourth line follows from the definitions in (\ref{eqn:snr_mimo_l}) and (\ref{eqn:r_code_l}).   

\subsection{Optimization Criteria}
\label{sec:opt_criteria}
All three criteria in Section \ref{sec:design_criteria}, namely, the minimization of the error volume $\operatorname{det}\left(\mathbf{R_e}^{mmse}\right)$ in (\ref{eqn:error_volume}), the minimization of the total \textit{nMSE} $\operatorname{Tr}\left\{\mathbf{nMSE}^{(g)}\right\}$ in (\ref{eqn:nmse_error_cov}), and the 
maximization of the mutual information given by 
$I\left(\mathbf{h}^{(g)};\mathbf{y}^{(g)}\right)$ in (\ref{eqn:mutual_inf_S_D}) lead to the following optimization problem: 
\begin{equation}
\mathbf{S}_{D,opt}^{(g)}=\underset{\mathbf{S}_D^{(g)}}
{\operatorname{argmin}} \quad
\operatorname{Tr}\left\{\left(
\boldsymbol{\mathcal{F}}^{(g)}+\mathbf{I}_{TD}\right)^{-1}\right\}
\label{eqn:nmse_S_D_opt}
\end{equation}
or
\begin{equation}
\mathbf{S}_{D,opt}^{(g)}=\underset{\mathbf{S}_D^{(g)}}
{\operatorname{argmax}} \quad
\operatorname{det}\left(\boldsymbol{\mathcal{F}}^{(g)}
+\mathbf{I}_{TD}\right)
\label{eqn:mutual_inf_det_S_D_opt}
\end{equation}
where
\begin{equation}
\boldsymbol{\mathcal{F}}^{(g)}
\triangleq 
\sum_{l=0}^{L_g-1}\mathbf{R}^{(g)}_{code}(l) \otimes 
\mathbf{SNR}^{(g)}_{mimo}(l).
\label{eqn:def_F_S_D_opt}
\end{equation}
\noindent Here, one can place an orthogonality constraint on the pre-beamformer such that $\left[\mathbf{S}_D^{(g)}\right]^H\mathbf{S}_D^{(g)}=\mathbf{I}_D$, which is desirable for random beamforming-type user scheduling \cite{nam14} and other two-stage beamforming based massive MIMO precoding \cite{kim15}.

It is observed that the optimization problem in (\ref{eqn:nmse_S_D_opt}) and (\ref{eqn:mutual_inf_det_S_D_opt}) has a non-trivial solution, yielding $\mathbf{S}_D^{(g)}$, for a given \textit{intra-group} pilot pattern $\left\{x_n^{(g_k)};\;-L_g+1 \leq n \leq T-1\right\}$ in (\ref{eqn:training_matrix_user_k}). 
The optimization metric $\boldsymbol{\mathcal{F}}^{(g)}$ in (\ref{eqn:def_F_S_D_opt}) depends on the temporal training pattern $\left\{x_n^{(g_k)}\right\}$ in addition to the spatial beam pattern $\mathbf{S}_D^{(g)}$ in a coupled manner. Therefore, the optimal pre-beamformer $\mathbf{S}_{D,opt}^{(g)}$ is expected to depend on the training pattern used in general. However, as it will be seen later, one can simplify the problem so that the pre-beamformer can be constructed independently from the training pattern. Before proposing a nearly optimal procedure to get $\mathbf{S}_D^{(g)}$, the following theorem is established. 

\underline{\textit{Theorem 1}}: For a given training pattern, constructing $\mathbf{X}^{(g)}$ in (\ref{eqn:r_code_l}), the two problems given in (\ref{eqn:nmse_S_D_opt}) and (\ref{eqn:mutual_inf_det_S_D_opt}) are equivalent.
 
\underline{\textit{Proof}}: First, the eigendecomposition of  
$\mathbf{R}^{(g)}_{code}(l)$ and $\mathbf{SNR}^{(g)}_{mimo}(l)$ matrices for $l=0,\ldots,L_g-1$ can be expressed as 
\begin{align}
\mathbf{R}^{(g)}_{code}(l)&=
\sum_{\left\{m \; \vert \beta_m^l > 0\right\}}
\beta_m^l \boldsymbol{\phi}_m^l 
\left[\boldsymbol{\phi}_m^l\right]^H 
\label{eqn:R_code_l_svd} \\
\mathbf{SNR}^{(g)}_{mimo}(l)&=
\boldsymbol{\Gamma}_l
\operatorname{diag}\left[\left\{\lambda_n^l\right\}_{n=1}^{D}\right]
\left(\boldsymbol{\Gamma}_l\right)^{-1}
\label{eqn:SNR_l_svd}
\end{align}
\noindent where $\boldsymbol{\Gamma}_l \triangleq \left[\boldsymbol{\gamma}_1^l \cdots 
\boldsymbol{\gamma}_D^l\right]_{D \times D}$ showing the $n^{th}$ dominant eigenvector $\boldsymbol{\gamma}_n^l$ in its $n^{th}$ column and $\lambda_n^l$ is defined as the corresponding eigenvalue of the $\mathbf{SNR}^{(g)}_{mimo}(l)$ matrix. In a similar fashion, $\boldsymbol{\phi}_m^l$ and $\beta_m^l$ are defined as the $m^{th}$ dominant eigenvector and eigenvalue of $\mathbf{R}^{(g)}_{code}(l)$ respectively. Then, the following Kronecker product rule $\left(\mathbf{R}^{(g)}_{code}(l) \otimes \mathbf{SNR}^{(g)}_{mimo}(l)\right) 
\left(\boldsymbol{\phi}_m^l \otimes \boldsymbol{\gamma}_n^l\right)=\left(\mathbf{R}^{(g)}_{code}(l) 
\boldsymbol{\phi}_m^l\right) \otimes \left(\mathbf{SNR}^{(g)}_{mimo}(l) \boldsymbol{\gamma}_n^l\right)=
\left(\beta_m^l\lambda_n^l\right)\left(\boldsymbol{\phi}_m^l \otimes \boldsymbol{\gamma}_n^l\right)$ implies that $\left\{\boldsymbol{\phi}_m^l \otimes \boldsymbol{\gamma}_n^l\right\}_{\forall m,n}$ is the set of eigenvectors for $\mathbf{R}^{(g)}_{code}(l) \otimes \mathbf{SNR}^{(g)}_{mimo}(l)$ with the corresponding set of eigenvalues $\left\{\beta_m^l\lambda_n^l\right\}_{\forall m,n}$ for $l=1,\ldots,L_g-1$. The matrix $\mathbf{R}^{(g)}_{code}(l) \otimes \mathbf{SNR}^{(g)}_{mimo}(l)$ is positive semi-definite, since $\beta_m^l\lambda_n^l$ values are non-negative (due to the positive semi-definiteness of $\mathbf{R}^{(g)}_{code}(l)$ and $\mathbf{SNR}^{(g)}_{mimo}(l)$). This implies that $\boldsymbol{\mathcal{F}}^{(g)}$, sum of $\mathbf{R}^{(g)}_{code}(l) \otimes \mathbf{SNR}^{(g)}_{mimo}(l)$ matrices in (\ref{eqn:def_F_S_D_opt}), is also positive semi-definite. Therefore, by defining the variables $\kappa_i,\;i=1,\ldots,TD$ as the eigenvalues of $\boldsymbol{\mathcal{F}}^{(g)}$, the problem of finding the optimal pre-beamformer $\mathbf{S}_D^{(g)}$, yielding $\kappa_i$s, is to minimize $\sum_{i=1}^{TD}\left(\kappa_i+1\right)^{-1}$ in (\ref{eqn:nmse_S_D_opt}) or to maximize $\prod_{i=1}^{TD}\left(\kappa_i+1\right)$ in (\ref{eqn:mutual_inf_det_S_D_opt}). It can be seen that the following optimization criteria are equivalent such that they result in the same optimal $\mathbf{S}_{D,opt}^{(g)}$, which yields the same eigenvalues $\kappa_i$s, since $\kappa_i \geq 0$ $\forall i$: $\operatorname{argmax}\prod_i\left(\kappa_i+1\right) \equiv \operatorname{argmin}\sum_i-\operatorname{log}\left(\kappa_i+1\right)
\equiv \operatorname{argmin}\sum_i\left(\kappa_i+1\right)^{-1}$. Finally, this argument establishes the equivalence between different criteria given in Section \ref{sec:design_criteria}. $\square$  

\subsection{Nearly Optimal Solution: Generalized Eigenvector Space}
\label{sec:nearly_opt_solution}
It is possible to simplify the optimization problem in (\ref{eqn:nmse_S_D_opt}) or (\ref{eqn:mutual_inf_det_S_D_opt}) where the optimal pre-beamformer depends on the training pattern in general. As mentioned earlier, \textit{channel sparsity} is pronounced in mm wave channels such that the AoA supports of each MPC is nearly non-overlapping in the \textit{angle-delay} plane \cite{swindlehurst16,adhikary14}. In addition to that, as the number of array elements $N$ increases, the eigenspaces of the covariance of each MPC tend to be nearly orthogonal. In light of this near-orthogonality assumption, the pre-beamformer of group $g$ can be constructed as     
\begin{equation}
\mathbf{S}_D^{(g)} \triangleq
\left[\begin{array}{cccc}
\mathbf{S}_D^{(g)}(0) & \mathbf{S}_D^{(g)}(1) & \cdots & \mathbf{S}_D^{(g)}(L_g-1)
\end{array}\right]_{N \times D}
\label{eqn:S_D_l_approx}
\end{equation}
\noindent where the $N \times d_l$ matrix $\mathbf{S}_D^{(g)}(l)$ can be seen as the sub-beamformer that allows $l^{th}$ resolvable MPC of group $g$ to pass while suppressing the \textit{inter-group interference} in the spatial domain, and $\sum_{l=0}^{L_g-1}d_l=D$. Due to the apparent near-orthogonality among the different MPCs (especially for mm wave frequencies), $\mathbf{S}_D^{(g)}(l)$ is also expected to reject each MPC of group $g$ other than the one at $l^{th}$ delay. Therefore, if the orthogonality among different MPCs is preserved after pre-beamforming, the dominant eigenvalues of the $D \times D$ $\mathbf{SNR}^{(g)}_{mimo}(l)$ matrix are the same as that of the $d_l \times d_l$ $\rho_l^{(g)}\left(\left[\mathbf{S}_D^{(g)}(l)\right]^H \mathbf{R}^{(g)}_{\boldsymbol{\eta}}
\mathbf{S}_D^{(g)}(l)\right)^{-1}\left(\left[\mathbf{S}_D^{(g)}(l)\right]^H 
\mathbf{R}_l^{(g)}\mathbf{S}_D^{(g)}(l)\right)$ matrix for $l=0,\ldots,L_g-1$, whereas the other eigenvalues of $\mathbf{SNR}^{(g)}_{mimo}(l)$ are nearly zero.
By using the definitions given in (\ref{eqn:R_code_l_svd}) and (\ref{eqn:SNR_l_svd}), the eigenspaces of each $\mathbf{SNR}^{(g)}_{mimo}(l)$ matrix (with dimensionality $\operatorname{rank}\left\{\mathbf{SNR}^{(g)}_{mimo}(l)\right\} \leq d_l)$ are mutually orthogonal in this case: $\boldsymbol{\Gamma}_{l_1}^H\boldsymbol{\Gamma}_{l_2}
\approx \mathbf{0}$ for $l_1 \neq l_2$.  
Then, the eigenspace of $\boldsymbol{\mathcal{F}}^{(g)}$ in (\ref{eqn:def_F_S_D_opt}) can be written as the \textit{orthogonal direct sum} of the eigenspaces of $\mathbf{R}^{(g)}_{code}(l) \otimes \mathbf{SNR}^{(g)}_{mimo}(l)$ matrices for $l=0,\ldots,L_g-1$, i.e., $\bigoplus_{l=0}^{L_g-1}\left\{\boldsymbol{\phi}_m^l \otimes \boldsymbol{\gamma}_n^l\right\}_
{\forall m,n}$ with the corresponding eigenvalues $\left\{\beta_m^l\lambda_n^l\right\}_{m,n}$. This leads to the following approximation of the optimization criterion in (\ref{eqn:nmse_S_D_opt}):
\begin{equation}
\operatorname{Tr}\left\{\mathbf{nMSE}^{(g)}\right\}=
\sum_{l=0}^{L_g-1}\sum_{m=1}^{R_l}\sum_{n=1}^{d_l}
\frac{1}{\beta_m^l \lambda_n^l+1} 
+\left(K_g\sum_{l=0}^{L_g-1}r_{g,l}
-\sum_{l=0}^{L_g-1}R_ld_l\right)
\label{eqn:nmse_error_cov_orthMPCs}
\end{equation}
\noindent where $R_l$ is the rank of $\mathbf{R}^{(g)}_{code}(l)$ matrix with $R_l \leq \operatorname{min}\left\{T,K_g\right\}$ from (\ref{eqn:r_code_l}). For a given dimension of the pre-beamformer $\mathbf{S}_D^{(g)}(l)$ in (\ref{eqn:S_D_l_approx}) with $\sum_l d_l=D$, and the training pattern determining $\beta_m^l$, it can be noted that the minimum value of the cost function $\operatorname{Tr}\left\{\mathbf{nMSE}^{(g)}\right\}$ in (\ref{eqn:nmse_error_cov_orthMPCs}), under the constraint that $\mathbf{S}_D^{(g)}$ is a full column rank matrix is achieved by the first $d_l$ dominant \textit{generalized eigenvectors} of $\mathbf{R}_l^{(g)}$ and $\mathbf{R}^{(g)}_{\boldsymbol{\eta}}$ from Appendix I. The minimum value of (\ref{eqn:nmse_error_cov_orthMPCs}) is attained by choosing $\lambda_n^l$ as the $n^{th}$ dominant \textit{generalized eigenvalue} of $\mathbf{R}_l^{(g)}$ and $\mathbf{R}^{(g)}_{\boldsymbol{\eta}}$ as noted in Appendix I. One can also determine the optimal $d_l$ values among the possible alternatives satisfying $\sum_l d_l=D$ by using the \textit{generalized eigenvalues} $\lambda_n^l$ minimizing (\ref{eqn:nmse_error_cov_orthMPCs}). It is observed that the pilot overhead is significantly reduced since even for $T=K_g$ (independent of $N$), $\operatorname{Tr}\left\{\mathbf{nMSE}^{(g)}\right\}$ approaches zero when $d_l=r_{g,l}$ and $R_l=K_g$.   



\section{Approximate Correlator type Estimators after Beamforming in High SNR Regime}
\label{sec:approximate_estimators}


In this section, a reduced rank correlator-type estimator in the spatio-temporal domain based only on the pre-beamforming matrix (designed by using only long-term channel statistical properties) is constructed. The key idea is to provide a reduced rank approximation of the \textit{Wiener} estimator in (\ref{eqn:h_eff_def}), which performs the optimum weighting of the decorrelated channel coefficients in the basis of the eigenvectors given by the columns of the KLT matrix (\ref{eqn:V_def}) defined in the spatio-temporal domain. This can be realized by applying the maximum likelihood (ML) estimator after a suitable subspace projection, which suppresses spatial interference while reducing dimensionality, thus yields optimum \textit{bias-variance tradeoff} \cite{utschick05,kaybook}.      

The ML (or zero-forcing) estimate, which is unbiased, achieves the Cramer-Rao bound (CRB) \cite{treesbook,kaybook}. The \textit{Wiener} estimator converges to the ML estimator as \textit{snr} approaches infinity after beamforming, i.e., $\frac{E_s}{N_0} \to \infty$. The ML estimator, based on the deterministic signal model in (\ref{eqn:y_eqn}), does not exploit spatial correlations in the multi-path channel vector, namely $\rho_l^{(g)}\mathbf{R}_l^{(g)}$. Particularly, ML variance approaches to infinity for low \textit{snr} \cite{kaybook}. This case is commonly encountered in mm wave channels where the \textit{snr} level at each antenna is expected to be very small before beamforming. Therefore, the principle of reducing the number of parameters to be estimated, without losing the intended part of the group $g$ signal, can be adopted again. Then, the reduced rank ML estimator as a post-processing stage can be constructed on a suitable subspace spanned by the columns of the pre-beamformer matrix $\mathbf{S}_D^{(g)}$. By this way, one can reduce the estimation error variance (or MSE) considerably at the expense of introduced bias when compared to the conventional (full dimensional) ML estimator, since the noise and interference subspace are switched off by the pre-beamforming in spatial domain. The nearly optimal pre-beamformer, constructed by the \textit{generalized eigenvector beamspace} (GEB) in Section \ref{sec:nearly_opt_solution}, is used here to reduce the dimensionality before applying the ML estimator. The GEB is an appropriate alternative for subspace projection, since it captures a significant portion of all MPCs in group $g$, while rejecting \textit{inter-group interference}.  

\subsection{High SNR Approximation for Angle Domain Estimator}
\label{sec:angle_domain_approx}
First, the reduced rank correlator-type approximation for the \textit{angle} domain RR-MMSE in (\ref{eqn:h_eq_mmse_approx}) is obtained. As mentioned in Section \ref{sec:angle_only_mmse}, it corresponds to the case where all MPCs of group $g$ are unified in the \textit{angular} domain, and the rank reduction is performed based on the spatial channel properties captured by 
$\mathbf{R}^{(g)}_{sum}$ in (\ref{eqn:Rsum_g}). If the GEB is used, the $\mathbf{SNR}^{total,(g)}_{mimo}$ matrix, obtained after pre-beamforming $\mathbf{S}_D^{(g)}$ in (\ref{eqn:snr_mimo_total}), is completely diagonalizable (as shown in Appendix I). In this case, the reduced rank ML estimate of the \textit{effective} channel in (\ref{eqn:h_eff_def}) can be obtained by letting $\frac{E_s}{N_0} \to \infty$ in (\ref{eqn:h_eq_mmse_approx}). This can be fulfilled by keeping $N_0$ fixed in (\ref{eqn:snr_mimo_total}), and allowing training power $E_s$ in (\ref{eqn:r_code_total}) to approach infinity. After following the mathematical steps provided in Appendix III, the RR-MMSE estimate of the \textit{effective} channel in (\ref{eqn:h_eq_mmse_approx}) can be approximated as  
\begin{equation}
\hat{\mathbf{h}}_{eff,2}^{(g)} \approx
\left\{ \begin{array}{cc}
\left\{\left(\left[\mathbf{X}^{(g)}\right]^H\mathbf{X}^{(g)}\right)^{-1}\left[\mathbf{X}^{(g)}\right]^H \otimes 
\left[\mathbf{S}_D^{(g)}\right]^H\right\}\mathbf{y} &
\qquad \textrm{ if } T \geq K_gL_g \\
\\
\left\{\left[\mathbf{X}^{(g)}\right]^H
\left(\mathbf{X}^{(g)}\left[\mathbf{X}^{(g)}\right]^H\right)^{-1} \otimes 
\left[\mathbf{S}_D^{(g)}\right]^H\right\}\mathbf{y} &
\qquad \textrm{ if } T < K_gL_g
\end{array}\right.
\label{eqn:h_eq_ml_approx}
\end{equation}
where $\mathbf{X}^{(g)}$ in (\ref{eqn:complete_training_matrix}) is assumed to be full column or row rank. 

%
%
%

\subsection{High SNR Approximation for Joint Angle-Delay Domain Estimator}
\label{sec:joint_angle_delay_domain_approx}
The joint \textit{angle-delay} domain RR-MMSE in (\ref{eqn:h_eq_mmse_est}) can also be approximated in a similar fashion by obtaining the ML estimate in reduced dimensional subspace. After pre-beamforming, it is assumed that the eigenspaces of each MPC of group $g$ are nearly orthogonal, an effect more strongly observed in mm wave channels especially for the case of a large number of antenna elements. In this case, the matrices $\mathbf{R}^{(g)}_{code}(l) \otimes \mathbf{SNR}^{(g)}_{mimo}(l)$ for $l=0,\ldots,L_g-1$ in (\ref{eqn:h_eq_mmse_est}) have orthogonal eigenspaces as explained in Section \ref{sec:nearly_opt_solution}, and $\mathbf{SNR}^{(g)}_{mimo}(l)$ matrices are completely diagonalizable with the use of GEB (see Appendix I). 
%
\subsubsection{Rank-1 Approximation}
\label{sec:joint_asymp_high_snr_rank1} 
By assuming that the rank of each MPC is one, i.e., $r_{g,l}=1$ for all $l$ in (\ref{eqn:mimo_MPC_covariance}), which is reasonable in the case of highly directional propagation, and using the pre-beamformer structure in (\ref{eqn:S_D_l_approx}), the following approximation for the RR-MMSE estimate of the \textit{effective} channel in (\ref{eqn:h_eq_mmse_est}) is obtained after some mathematical steps given in Appendix III:
\begin{equation}
\hat{\mathbf{h}}_{eff}^{(g)} \approx
\sum_{l=0}^{L_g-1}
\left( \underbrace{\operatorname{pinv}\left\{\mathbf{X}^{(g)} 
\left[\mathbf{I}_{K_g} \otimes \mathbf{E}_{L_g,l}\right]\right\}}_
{\textrm{(temporal) correlator}}
\otimes \underbrace{\left[\mathbf{S}_D^{(g)}\mathbf{E}_{D,l}\right]^H}_
{\textrm{pre-beamformer}}\right)\mathbf{y}
\label{eqn:h_eq_ml_approx_orthMPCs}
\end{equation}
\noindent where $\operatorname{pinv}\left\{\;\right\}$ is a generalized operation known as \textit{Moore-Penrose pseudoinverse}, used to obtain the inverse of singular or non-square matrices. In (\ref{eqn:h_eq_ml_approx_orthMPCs}), $\mathbf{E}_{D,l}$ is defined as an $D \times D$ elementary diagonal matrix where all the entries except the $\left(l+1\right)^{th}$ diagonal one are zero, and the dimension of the pre-beamformer is set as $D=L_g$. Here, $\mathbf{S}_D^{(g)}\mathbf{E}_{D,l}$ is steered towards the AoA of the $l^{th}$ MPC of group $g$  while rejecting other MPCs of groups $g$ and \textit{inter-group interference}. After the beamspace processing, the temporal processing, in the form of correlator, is applied in order to differentiate between the MPCs of all group $g$ users at $l^{th}$ delay, and to combat with other interfering sources (with overlapping AoA support) by simply placing temporal finger on the $l^{th}$ temporal diversity path for all $K_g$ \textit{intra-group} users.              

%
%

\subsubsection{Approximation for General-Rank Signal Models}
\label{sec:joint_asymp_high_snr_general_rank}
The approximate spatio-temporal correlator in (\ref{eqn:h_eq_ml_approx_orthMPCs}) can be extended to the more general case, where the rank of each resolvable MPC covariance is greater than one, and there exists significant overlap among some of the MPCs in the angular domain. In this case, one can simplify the problem by partitioning the MPCs of group $g$ into groups in the angular domain such that some of the MPCs at a specific delay, having approximately similar eigenspaces (common AoA support), are placed into the same group. The key idea is to construct resolvable MPC groups whose AoA supports are nearly orthogonal in the angular domain so that the reduced rank ML estimator for each group of resolvable MPCs can be realized separately. By assuming that the eigenspaces of each $\mathbf{SNR}^{(g)}_{mimo}(l)$ matrix in (\ref{eqn:h_eq_mmse_est}) are mutually orthogonal as explained in Section \ref{sec:nearly_opt_solution}, and letting $\frac{E_s}{N_0} \to \infty$, the following general approximation for (\ref{eqn:h_eq_mmse_est}) is obtained after carrying out similar mathematical steps to (\ref{eqn:h_eq_ml_approx_orthMPCs}):  
\begin{equation}
\hat{\mathbf{h}}_{eff}^{(g)} \approx 
\sum_{l=0}^{\mathcal{MPC}-1}
\left( \underbrace{\operatorname{pinv}\left\{\mathbf{X}^{(g)} 
\left[\mathbf{I}_{K_g} \otimes \sum_{m \in \mathfrak{L}_l} \mathbf{E}_{L_g,m}\right]\right\}}_{\textrm{(temporal) correlator}} \otimes \underbrace{\left[\mathbf{S}_D^{(g)}\sum_{n \in \mathfrak{D}_l}\mathbf{E}_{D,n}\right]^H}_{\textrm{pre-beamformer}}\right)\mathbf{y}
\label{eqn:h_eq_ml_approx_orthMPCs_general}
\end{equation}
\noindent where $\mathcal{MPC}$ is the total number of resolvable MPC groups in $g$ having nearly non-overlapping AoA support, $\mathfrak{L}_l$ is the set of non-zero (temporal) delays belonging to the $l^{th}$ resolvable multi-path group (non-overlapping AoA support) in the angular domain, and $\sum_{l=0}^{\mathcal{MPC}-1}\vert\mathfrak{L}_l\vert=L_g$. In (\ref{eqn:h_eq_ml_approx_orthMPCs_general}), the set $\mathfrak{D}_l$ is defined as $\mathfrak{D}_l \triangleq \left\{ n \in \mathbb{Z}^+ \vert \sum_{m=0}^{l-1}d_m < n \leq \sum_{m=0}^{l}d_m \right\}$ for $l>0$, and $\mathfrak{D}_l \triangleq \left\{ n \in \mathbb{Z}^+ \vert 0 < n \leq d_0 \right\}$ for $l=0$ where 
$\vert\mathfrak{D}_l\vert=d_l$ and $\sum_{l=0}^{\mathcal{MPC}-1}d_l=D$. Here, 
$\mathfrak{D}_l$ shows the column indices of the pre-beamformer matrix $\mathbf{S}_D^{(g)}$ in (\ref{eqn:S_D_l_approx}) allowed to pass the $l^{th}$ resolvable MPC of group $g$. In (\ref{eqn:h_eq_ml_approx_orthMPCs_general}), $\mathbf{S}_D^{(g)}$ is constructed as in (\ref{eqn:S_D_l_approx}), and 
$\mathbf{S}_D^{(g)}\sum_{n \in \mathfrak{D}_l}\mathbf{E}_{D,n}$, whose non-zero columns equal to that of $\mathbf{S}_D^{(g)}(l)$ in (\ref{eqn:S_D_l_approx}), can be thought as the $N \times D$ beamformer matrix obtained by replacing all sub-matrices in (\ref{eqn:S_D_l_approx}) with zero matrix except $\mathbf{S}_D^{(g)}(l)$. The sub-beamformer matrix $\mathbf{S}_D^{(g)}(l)$ is designated to reject other MPCs of group $g$ in addition to the \textit{inter-group interference}, and to capture a significant portion of the $l^{th}$ resolvable MPC (in a similar way to the one explained in Section \ref{sec:nearly_opt_solution}).     

In (\ref{eqn:h_eq_ml_approx_orthMPCs_general}), $\operatorname{pinv}\left\{\;\right\}$ operation can be seen as the temporal correlator preceded by the pre-beamformer. It performs the task of \textit{Least Square} (LS) type estimation of reduced dimensional channels corresponding to the $l^{th}$ MPC in group $g$. The form of (\ref{eqn:h_eq_ml_approx_orthMPCs_general}) appears as the \textit{decoupled} spatio-temporal processing where spatial pre-beamforming and temporal (Rake-type) correlator are applied in a successive manner. This further simplifies the RR-MMSE estimator in (\ref{eqn:h_eq_mmse_est}). For $L_g=1$, $D=N$ (no dimension reduction), and $\mathbf{S}_D^{(g)}=\mathbf{I}_N$, i.e., the spatial covariance structure of the MIMO channel is not exploited, the approximate estimator in (\ref{eqn:h_eq_ml_approx_orthMPCs_general}) reduces to the conventional LS type CSI acquisition technique, well-known in the literature \cite{swindlehurst14,ashikhmin11}, which relies on correlating the received signal with the known pilot sequence and suffering from pilot contamination, whereas with the use of pre-beamformer in (\ref{eqn:h_eq_ml_approx_orthMPCs_general}), the \textit{inter-group} interfering users leading to pilot interference are mitigated in the spatial domain. Moreover, the estimator in (\ref{eqn:h_eq_ml_approx_orthMPCs_general}) does not necessitate the a-priori power profile given by KLT in the \textit{angle-delay} domain.          
 
\section{Numerical Results and Discussion}
\label{sec:Sim_Results}
In this section, we provide some numerical results to evaluate the performance of the reduced rank channel estimators and examine the efficiency of the GEB in Section \ref{sec:nearly_opt_solution} for the reduced dimensional processing. Throughout the demonstrations, we consider a massive MIMO system with uplink training in TDD mode where a BS is equipped with a uniform linear array (ULA) of $N=100$ antenna elements along the y-axis%
\footnote{Although the system model and the proposed estimators are valid for an arbitrary array structure in this work, ULA is considered for ease of exposition only.}, 
and each of $K$ users has a single receive antenna. 

In the studied scenario, $K$ users were clustered into eight groups ($G=8$), and each UT is assumed to be located at a specific azimuth angle $\theta$ along the ring centered at the origin in x-y plane. Here, we assume users come in groups, either by nature or by the application of proper \textit{user grouping} algorithms in \cite{nam14,adhikary14}, which are out of scope of this work. The channel covariance matrix of each group is specified with the center azimuth angle $\theta$ (AoA), and can be calculated in a similar way to the ones in \cite{adhikary13,kim15}. In the simulations, our focus is on the channel estimation accuracy of the intended group 
$g$ with $3$ MPCs, i.e., $L_g=3$. The first two MPCs of group $g$ stem from a azimuth angular sector $[-1^\circ,1^\circ]$ for delays at $l=0,1$, and the angular sector of the last MPC at $l=2$ of $g$ is given as $[5^\circ,7^\circ]$ in azimuth. We assume two users served simultaneously for group $g$, i.e., $K_g=2$. Each of the other $7$ groups (interfering with the intended one) consists of three users, i.e., $K_{g'}=3,\;g' \neq g$ and these users have $3$ MPCs whose angular sectors have same supports of AoA ($L_{g'}=3,\;g' \neq g$) given by 
$[-29 , -26]$, $[-21 , -19]$, $[-12 , -9]$, $[-5.5 , -3.5]$, 
$[9.5 , 12.5]$, $[15 , 17]$, $[24 , 27]$ in azimuth respectively. The channel vector for each user is independently generated according to the model (\ref{eqn:KLT_expansion}). The noise power is set as $N_0=1$ so that all dB power values are relative to $1$. In TDD mode, \textit{inter-group} users do not need to be synchronized, and even are allowed to use the same sequences during uplink training mode. \textit{Intra-group} users (of the intended group) use non-orthogonal training waveforms composed of $6$ symbols ($T=6$), and these are obtained by truncating length-63 \textit{Kasami} codes \cite{goldsmithbook} by simply choosing the first $T$ symbols of last $K_g$ \textit{Kasami} sequences without any optimization%
\footnote{There are more efficient approaches (other than the truncation of Kasami codes) yielding waveforms with better cross- and auto-correlation properties and minimizing (\ref{eqn:nmse_S_D_opt}), but training optimization is beyond the scope of this study}.
Then, the training matrix in (\ref{eqn:complete_training_matrix}) can be constructed to be exploited by the BS during the CSI acquisition period. 

The trace of the estimation error covariance matrix (for the extended channel vector of group $g$ users in (\ref{eqn:vector_def2}) given by $\mathbf{R_e} \triangleq \mathbb{E}\left\{\left(\mathbf{h}^{(g)}-
\hat{\mathbf{h}}^{(g)}\right)
\left(\mathbf{h}^{(g)}-\hat{\mathbf{h}}^{(g)}\right)^H\right\}$ is evaluated to compare the performance of different estimators. Here, the channel estimates in the original space and in the reduced dimensional subspace after pre-beamforming are defined as 
$\hat{\mathbf{h}}^{(g)} \triangleq \left(\mathbf{W}^{(g)}\right)^H \mathbf{y}$ and $\hat{\mathbf{h}}^{(g)}_{eff} \triangleq \left(\mathbf{W}^{(g)}_{eff}\right)^H \mathbf{y}$ respectively. For these arbitrary linear estimators, the error covariance can be calculated by using the $\mathbf{R_e}^{mmse}$ in (\ref{eqn:mmse_error_cov}) (achieved by the reduced rank \textit{Wiener} filter (\ref{eqn:h_mmse_est})):
\begin{equation}
\mathbf{R_e} = \mathbf{R_e}^{mmse} + \left(\mathbf{W}^{(g)}-
\mathbf{W}^{(g)}_{mmse}\right)^H\mathbf{R}_{\mathbf{y}}
\left(\mathbf{W}^{(g)}-\mathbf{W}^{(g)}_{mmse}\right) 
\label{eqn:error_cov}
\end{equation}
\noindent where $\mathbf{W}^{(g)}_{mmse} \triangleq \boldsymbol{\Upsilon}_{U}^{(g)}\left(\mathbf{W}_{mmse,D}^{(g)}\right)^H
\left(\boldsymbol{\Upsilon}_S^{(g)}\right)^H$ in (\ref{eqn:MMSE_1}), and $\mathbf{W}^{(g)}$ is an $(NT) \times (NK_gL_g)$ arbitrary filter. In a similar manner, the error covariance matrix of the \textit{effective} channel estimate (\ref{eqn:h_eq_mmse_est}) can be calculated in the reduced dimensional subspace. The covariance matrix of the \textit{inter-group interference} is evaluated by (\ref{eqn:InterGroupInterference}) when the angular sector of each group is provided.  

In this study, we compare the performance of dimension reduction based on the GEB (shown to be nearly optimal under some realistic assumptions) with that of the conventional subspace composed of the first $D$ dominant eigenvectors of $\mathbf{R}^{(g)}_{sum}$ in (\ref{eqn:Rsum_g}). We call this conventional beamspace as discrete Fourier transform (DFT) beamspace because the eigenvectors of the spatial correlation matrix of the ULA channel are well approximated by the columns of the $N \times N$ unitary DFT matrix whose indices correspond to the support of the Fourier transform of the spatial correlation function (owing to the Szeg\H{o}'s asymptotic theory) \cite{nam14} depending on the angular sector of group $g$ UTs. This conventional beamspace is known to be information preserving for the spatially white interference case, and thus, is widely used in practical \textit{hybrid beamforming} applications, where the beamforming in the RF analog domain can be implemented by simple phase shifters \cite{liu14}. 

\begin{figure}[htbp]
\centering
\epsfig{file=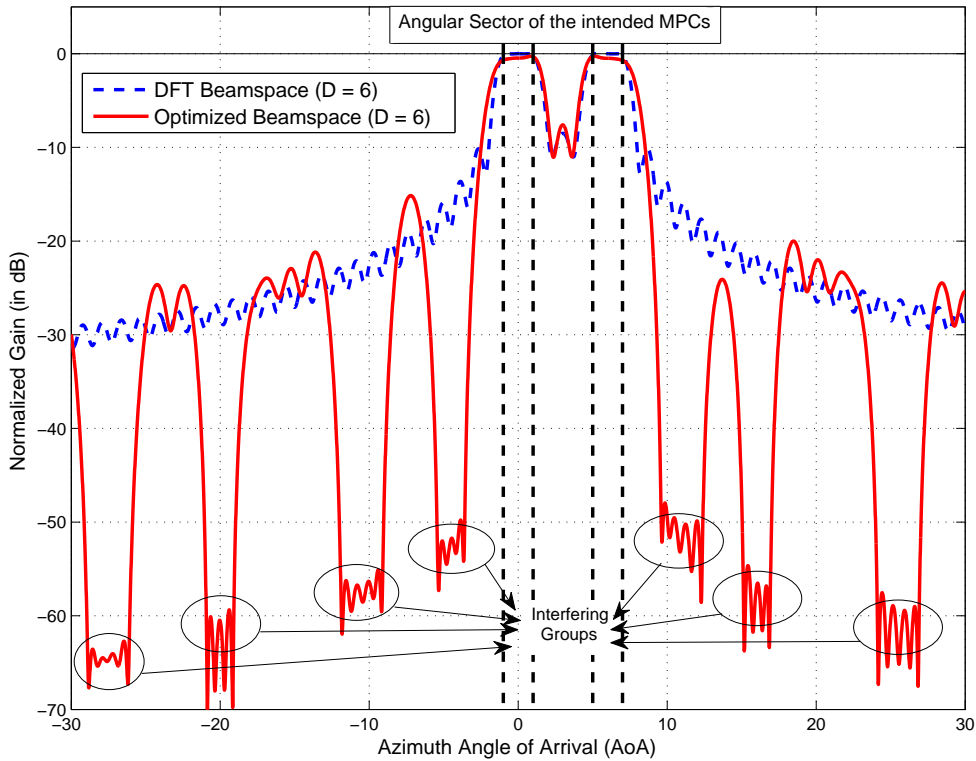,width=0.75\textwidth}
\caption{Beam pattern of different pre-beamformers}  
\label{fig:Pattern_Dim6_equalINR_SNR40}
\end{figure}

In Figure \ref{fig:Pattern_Dim6_equalINR_SNR40}, the beam patterns created by the GEB and DFT beamspaces are depicted for $D=6$ at $snr=30$   
(dB). The GEB is designed based on the AoA support of the intended group $g$ for $l=0,1,2$ while taking the angular locations of the interfering groups into account. The \textit{inter-group} users' signals are assumed to have the same power level with that of the intended group. As can be seen from the figure, the GEB tries to create deep nulls at the angular locations of interfering UTs, whereas the conventional pre-beamformer only tries to maximize the captured power of the intended group MPCs for a given dimension. It is expected that as the number of BS antennas increases, the eigenspaces of each group are approximately orthogonal. However, the number of transmit antennas is finite in practice, and there always exists some overlap among the virtual angular sectors of each group which leads into a leakage to the intended group signal. Therefore, as it will be shown later, the accuracy of the channel estimation realized on the reduced dimensional subspace, spanned by the conventional DFT beamspace, is considerably lost due to the residual \textit{inter-group interference} after pre-beamforming. On the other hand, the GEB suppresses the \textit{inter-group interference} while allowing the MPCs of the intended group to pass with a negligible distortion so that the subsequent processing in reduced dimensions, here the instantaneous CSI estimation, can be carried out as accurately as possible.     

\begin{figure}[htbp]
\centering
\epsfig{file=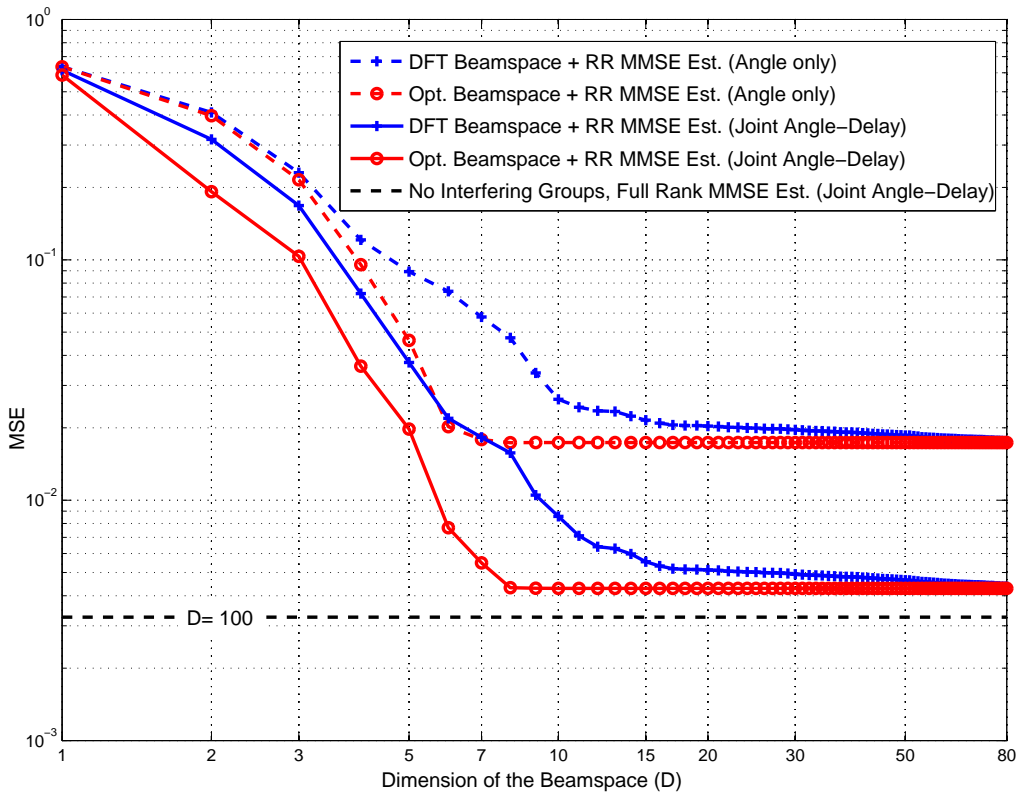,width=0.75\textwidth}
\caption{MSE values of RR-MMSE estimator versus dimension of the pre-beamformer}  
\label{fig:MSE_vs_Dim_SNR30_equalINR}
\end{figure}

In Figure \ref{fig:MSE_vs_Dim_SNR30_equalINR}, the average mean square error (MSE) values given by 
$\operatorname{Tr}{\left\{\mathbf{R_e}\right\}}/K_g$ as a function of the dimension of the spatial domain pre-beamformer ($D$) are depicted for both joint \textit{angle-delay} domain and \textit{angle} domain RR-MMSE estimators given in (\ref{eqn:h_mmse_est}) at $snr=30$ dB. For joint \textit{angle-delay}, the exact knowledge of covariance for each MPC is used, whereas for \textit{angle} domain, a common angular region (obtained by the unification of each delay) is assumed for each MPC (of group $g$) and used instead of $\mathbf{R}_l^{(g)}$ in (\ref{eqn:h_mmse_est}). Also, the performance of the full dimensional \textit{Wiener} estimator ($D=N$) is demonstrated when there are no interfering groups. It is clear that \textit{angle} domain RR-MMSE estimator is inferior to joint \textit{angle-delay} domain estimator due to the inefficient use of the training and noise enhancement. Moreover, it is seen that there is a remarkable performance gap between the performances of RR-MMSE estimators based on two different pre-beamformers (the GEB and the conventional one) especially at lower dimensions. Also, it can be concluded that the RR-MMSE estimator based on the GEB achieves a very close performance to that of the full dimensional estimator even for $D=7$ (for group $g$), that is roughly $15$ fold dimension reduction. On the other hand, with the conventional beamspace, in spite of the optimal \textit{Wiener filtering} after dimension reduction, the MSE performance is not satisfactory for dimensions below $14$.   

\begin{figure}[htbp]
\centering
\epsfig{file=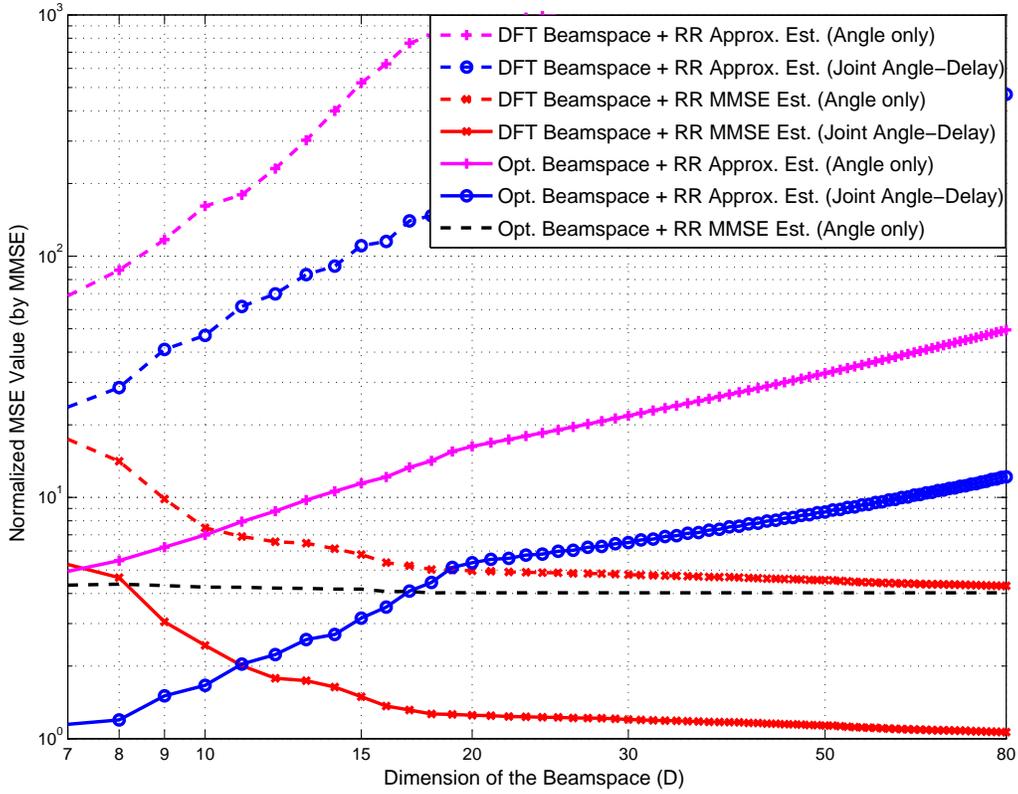,width=0.75\textwidth}
\caption{Normalized MSE values for different type of reduced rank CSI estimators versus dimension of the pre-beamformer}  
\label{fig:effectiveMSE_vs_Dim_SNR30_equalINR}
\end{figure}

In Figure \ref{fig:effectiveMSE_vs_Dim_SNR30_equalINR}, by adopting the same settings used to obtain in Figure \ref{fig:MSE_vs_Dim_SNR30_equalINR}, the MSE performance for different estimators of the \textit{effective} channel in $D$-dimensional spatial subspace are shown. The performance of joint \textit{angle-delay} RR-MMSE estimator based on the GEB in (\ref{eqn:h_eq_mmse_est}) is used as the performance benchmark. The MSE achieved by different \textit{effective} channel estimators are normalized by this benchmark value for each dimension, and these relative MSE values are given as a function of the dimension $D$ (starting at $7$). Also, the spatio-temporal correlator type estimators in (\ref{eqn:h_eq_ml_approx}) and (\ref{eqn:h_eq_ml_approx_orthMPCs_general}), obtained after the high \textit{snr} approximation of reduced rank \textit{Wiener filter}, are depicted. It is seen that relative performance of these approximate correlator type estimators degrades as the dimension increases. This degradation is expected, since LS type estimation does not exploit spatial correlations in the channel coefficients, and does not apply optimum spatial weights when compared to MMSE filtering. In this case, adding extra dimensions beyond $7$ leads to noise enhancement, since the noise subspace is not switched off properly with increasing dimensions, and starts to contaminate the \textit{effective} channel estimates. In addition to that, the approximate estimator is more sensitive to which pre-beamformer is utilized such that there is a remarkable performance gap between the GEB and the conventional beamspaces when the approximate estimator is realized for both joint \textit{angle-delay} and
\textit{angle} domain. This is reasonable since the GEB rejects the interference subspace properly while reducing dimensionality before applying the reduced rank LS estimator. The proposed approximate estimator appears to be so effective that the benchmark performance is attained without using the exact knowledge of the spatial correlation matrices at significantly reduced complexity (dimension). Moreover, both the RR-MMSE and the correlator type estimators, based on the sparsity information in the joint \textit{angle-delay} domain, are able to remove the pilot contamination by mitigating the \textit{inter-group interference} without the need of any pilot coordination.       

\begin{figure}[htbp]
\centering
\epsfig{file=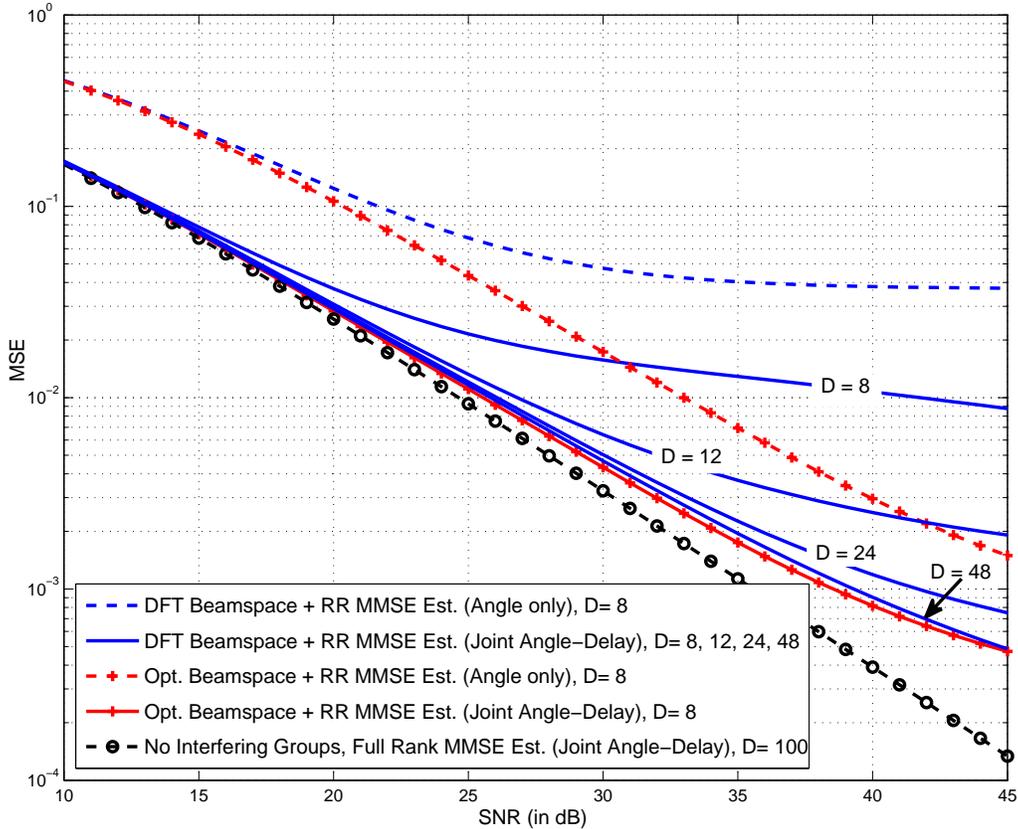,width=0.75\textwidth}
\caption{MSE values for the RR-MMSE estimators based on different type of pre-beamformers versus \textit{snr}}  
\label{fig:MSE_vs_SNR_Dim8_12_24_48_equalINR}
\end{figure}

In Figure (\ref{fig:MSE_vs_SNR_Dim8_12_24_48_equalINR}), the MSE values of the RR-MMSE estimators in (\ref{eqn:h_mmse_est}) as a function of the \textit{snr} (after beamforming) are depicted for various dimension values. It is observed that processing based on the DFT beamspace needs much larger dimensions to obtain the same accuracy level with that of the estimators based on the GEB. 

\begin{figure}[htbp]
\centering
\epsfig{file=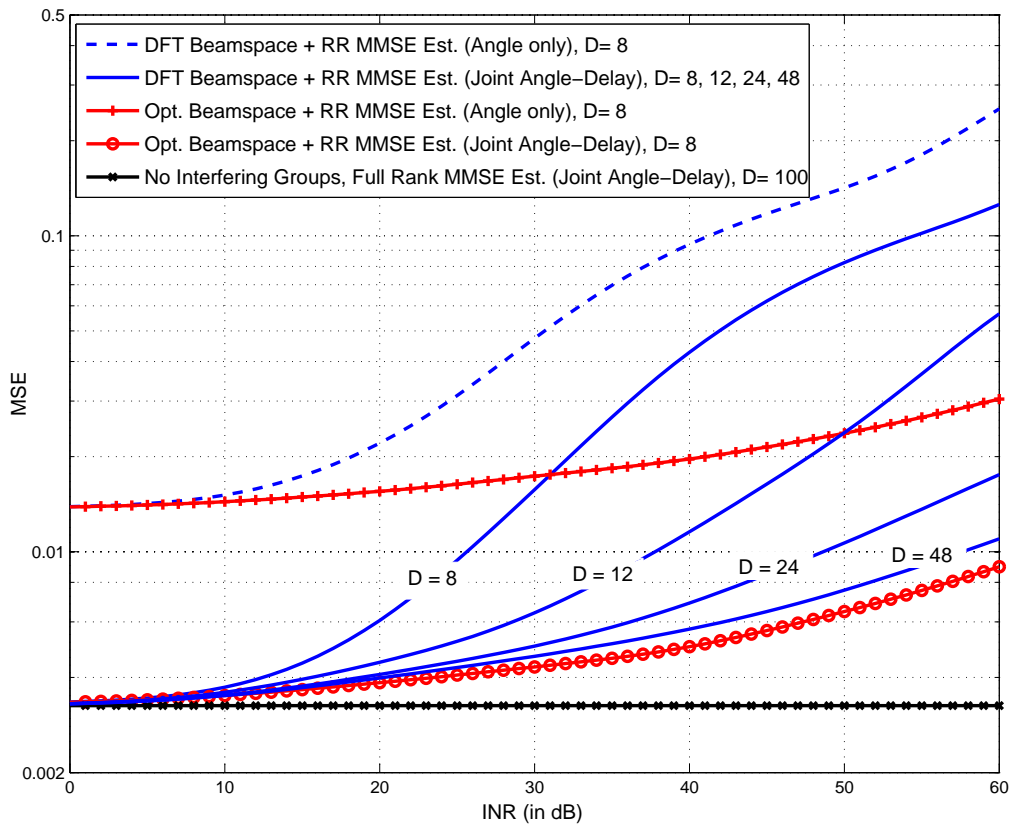,width=0.75\textwidth}
\caption{MSE values for the RR-MMSE estimators based on different type of pre-beamformers versus \textit{inr}}  
\label{fig:MSE_vs_INR_Dim8_12_24_48_SNR30}
\vskip -1em
\end{figure}

In Figure \ref{fig:MSE_vs_INR_Dim8_12_24_48_SNR30}, the effect of the interference level, i.e., the interference-to-noise ratio (\textit{inr}), on the MSE values achieved by the RR-MMSE is depicted. It is observed that the performance gap (from the full dimensional operation with no interference) increases drastically as the \textit{inr} increases for the conventional beamspace at lower dimensions. On the other hand, graceful degradation is observed for the GEB at $D=8$. This shows that the conventional subspace, constructed without the statistical knowledge of the interfering sources, needs much larger dimension in order to suppress the \textit{inter-group interference} properly especially when the received signal strength of different group UTs may differ significantly depending on their distance to the BS (\textit{near-far} effect).      

\begin{figure}[htbp]
\centering
\epsfig{file=
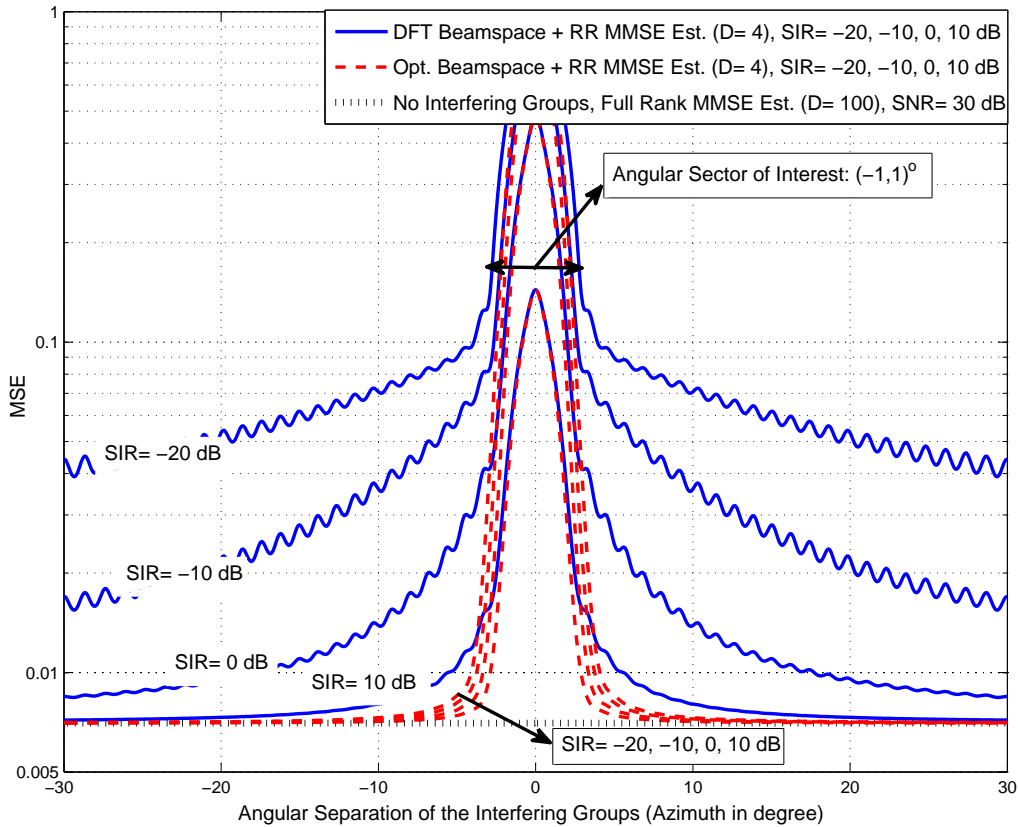,width=0.75\textwidth}
\caption{MSE values for the RR-MMSE estimators based on different type of pre-beamformers versus the angular separation in azimuth}  
\label{fig:MSE_vs_AngularSeparation_Dim4_SNR30}
\vskip -1em
\end{figure}

As a final comparison, the effect of different interference levels on the MSE values are investigated when the nearly optimal GEB and conventional subspace based dimension reduction technique are considered at $snr=30$ dB and $D=4$. It is assumed that there are only two groups interfering with each other; the intended one has angular sector $[-1^\circ,1^\circ]$ with $2$ MPCs, and the other has a similar AS with that of the intended one with a varying AoA. Figure \ref{fig:MSE_vs_AngularSeparation_Dim4_SNR30} depicts the MSE values achieved by RR-MMSE for various \textit{inr} values as a function of the angular separation between the intended group and the interfering one. As can be seen from the beam pattern of two different pre-beamformers in Figure \ref{fig:Pattern_Dim6_equalINR_SNR40}, the conventional one is not able to suppress \textit{inter-group interference} as much as the GEB does. Thus, for smaller signal-to-interference ratio (\textit{sir}) values, i.e., when the \textit{near-far} effect is more apparent, the residual interference after pre-beamforming still affects the performance of the RR-MMSE estimator dramatically even for larger angular separation when conventional beamspace is used. However, the GEB perfectly differentiates between groups while reducing the dimension before fine CSI acquisition so that the RR-MMSE attains the MSE of full dimensional filter even for small angular separation at significantly reduced complexity and pilot overhead.     

\section{Conclusions}
\label{sec:Conclusion}
The processing of the signals with very large dimensionality, the pilot interference, and the pilot overhead are thought to be limiting factors for an accurate channel acquisition and throughput of massive MIMO transmission in mm wave especially in high mobility or in applications requiring low latency and short-packet duration. In this paper, a general framework for the reduced-dimensional massive MIMO channel estimation problem was established based on the statistical \textit{user-grouping} (in the JSDM scheme) when the statistical pre-beamformer was designed to reduce dimensionality and pilot overhead while mitigating \textit{inter-group interference} leading to pilot contamination (due to inner or outer cell users). 

The main contributions of the paper are summarized as follows. First, the RR-MMSE channel estimator, based on generic subspace projection and the second order statistics, was presented for the first time when the SC transmission in TDD mode for wideband multi-user spatially correlated MIMO systems was considered. It can be interpreted as the reduced rank approximation of the optimal spatio-temporal \textit{Wiener filter} by using two different generic transform basis sets; namely, the dimension-reducing subspace projection (pre-beamformer), and the KLT characterizing channel sparsity in joint \textit{angle-delay} domain. Second, we examined the dimension reduction problem from three different viewpoints related to the instantaneous CSI estimation accuracy. The goal was to find a good beamspace (subspace in spatial domain) on which the reduced dimensional channel estimation can be fulfilled as accurately as possible. The adopted criteria of the problem resulted in three equivalent optimization problems yielding the same optimal pre-beamformer. After some reasonable and practical approximations, the generalized eigenvector beamspace was shown to be a nearly optimal pre-beamformer (when the eigenspaces of different resolvable MPCs are assumed to be nearly orthogonal). Finally, the reduced rank correlator type estimator in the spatio-temporal domain was proposed by applying LS estimation after a subspace projection (pre-beamformer), which suppresses spatial interference while reducing dimensionality. The structure of the estimator appeared as the decoupled spatio-temporal processing where
spatial pre-beamforming and temporal correlator were applied in a successive manner separately. Different from the conventional LS estimators, the proposed technique here, is a kind of covariance-aware LS estimator achieving the optimum bias-variance tradeoff. The proposed estimators show very close performance to that of the full dimensional \textit{Wiener} estimator at significantly reduced complexity. Moreover, they demonstrate remarkable robustness to the pilot contamination with a significant reduction in pilot overhead with the help of properly designed pre-beamformer which mitigates the inter-group interfering
users leading to pilot interference in the spatial domain.  

To sum up, in this paper, we provide a general description for massive MIMO transmission employing SC in frequency-selective fading. The proposed covariance-based reduced rank estimators together with the beamformer design here, confirm, compare, and complement many previous works, where the pilot interference due to the use of non-orthogonal pilots among the intra- or
inter-cell users persists, by changing several system and model parameters.        
Adaptive learning of \textit{long-term parameters} (such as AoA supports and delays), and adaptive subspace construction (or tracking) with \textit{user-grouping}, under the proposed beamformer design and CSI acquisition framework, can be topics of future studies. The effect of inaccurate second order statistical information or mismatches (related to the joint \textit{angle-delay} power profile or low-rank spatial channel covariance matrices) on the performance of reduced rank channel estimators' can be investigated.    

\section*{Appendix I: Properties of the $\mathbf{SNR}$ Matrix}
\label{sec:Appendix_1}
We can list important properties of the $\mathbf{SNR}^{(g)}_{mimo}(l)$ matrix as follows:
\begin{enumerate}
\item[1.] $\mathbf{SNR}^{(g)}_{mimo}(l)$ matrix in 
(\ref{eqn:snr_mimo_l}) and $\mathbf{SNR}^{total,(g)}_{mimo}$ matrix in (\ref{eqn:snr_mimo_total}) are positive semi-definite, since
$\left(\left[\mathbf{S}_D^{(g)}\right]^H 
\mathbf{R}^{(g)}_{\boldsymbol{\eta}}\mathbf{S}_D^{(g)}\right)$, 
$\left(\left[\mathbf{S}_D^{(g)}\right]^H \mathbf{R}_l^{(g)}
\mathbf{S}_D^{(g)}\right)$, and
$\left(\left[\mathbf{S}_D^{(g)}\right]^H \mathbf{R}^{(g)}_{sum}
\mathbf{S}_D^{(g)}\right)$
are positive semi-definite, and so is their
inverse and their multiplication.
\item[2.] Generalized eigenvectors of $\mathbf{R}_l^{(g)}$ 
(or $\mathbf{R}^{(g)}_{sum}$) and
$\mathbf{R}^{(g)}_{\boldsymbol{\eta}}$ matrices diagonalize the 
$\mathbf{SNR}^{(g)}_{mimo}(l)$ (or $\mathbf{SNR}^{total,(g)}_{mimo}$) matrix. Stated differently, if $\mathbf{R}_l^{(g)}\mathbf{v}_n = 
\mathbf{R}^{(g)}_{\boldsymbol{\eta}}\mathbf{v}_n
\lambda_n^l$, where $\lambda_n^l$ and $\mathbf{v}_n$ are the $n^{th}$
largest generalized eigenvalue ($\lambda_1^l \geq \lambda_2^l \geq
\ldots > \lambda_N^l$) and its associated eigenvector, then a basis
for $N$ dimensional space can be written as
\begin{equation}
\mathbf{S}_N^{(g)}=\left[ \mathbf{v}_1\, \mathbf{v}_2\, \ldots \,
\mathbf{v}_N\right].
\label{eqn:gebasis}
\end{equation}
\noindent The generalized eigenvectors of symmetric matrices have the
property of being $\mathbf{R}^{(g)}_{\boldsymbol{\eta}}$ orthogonal,
$\mathbf{v}^H_k\mathbf{R}^{(g)}_{\boldsymbol{\eta}}\mathbf{v}_n = 0$, $n \neq k$, and $\mathbf{v}^H_n\mathbf{R}^{(g)}_{\boldsymbol{\eta}}\mathbf{v}_n = 1$, and can be $\mathbf{R}^{(g)}_{\boldsymbol{\eta}}$-\textit{orthonormalized} as follows.
When $\mathbf{S}_N^{(g)}$ given in (\ref{eqn:gebasis}) is inserted in
(\ref{eqn:snr_mimo_l}), the $\mathbf{SNR}^{(g)}_{mimo}(l)$ reduces to a diagonal matrix with the generalized eigenvalues on its diagonal.
\item[3.] An alternative representation for the vectors of $N$ dimensions, that is another basis for the subspace spanned by the columns of $\mathbf{S}_D^{(g)}$ in (\ref{eqn:gebasis}), results in a \textit{similarity transformation} for $\mathbf{SNR}^{(g)}_{mimo}(l)$ matrix. Stated differently, if $\mathbf{S}_D^{(g)}$ is
replaced with $\mathbf{S}_D^{(g)} \mathbf{T}$ in (\ref{eqn:snr_mimo_l}) where $\mathbf{T}$ is a $D \times D$ invertible matrix, 
$\mathbf{SNR}^{(g)}_{mimo}(l)$ matrix becomes
$\mathbf{T}^{-1} \mathbf{SNR}^{(g)}_{mimo}(l) \mathbf{T}$.
\item[4.] The cost/reward functions such as trace and determinant, which are invariant to the basis representation, remain invariant when
applied to the $\mathbf{SNR}^{(g)}_{mimo}(l)$ matrix. Any other cost function depending solely on the eigenvalues of the
$\mathbf{SNR}^{(g)}_{mimo}(l)$ matrix has the same property. For such functions, we may consider that basis vectors, spanning the subspace, are $\mathbf{R}^{(g)}_{\boldsymbol{\eta}}$-\textit{orthonormalized} without any loss of generality.
\item[5.] If we consider $\operatorname{Tr}\left\{
\left(\mathbf{R}^{(g)}_{code}(l) \otimes
\mathbf{SNR}^{(g)}_{mimo}(l)+\mathbf{I}_{TD}\right)^{-1}\right\}$ as the cost function, the minimum cost that can be achieved is
$\sum_{m=1}^T\sum_{n=1}^D\left(1+\beta_m^l\lambda_n^l\right)$ where 
$\beta_m^l$ are the non-negative eigenvalues of $\mathbf{R}^{(g)}_{code}(l)$ in (\ref{eqn:r_code_l}), and $\lambda_n^l$ is the non-negative valued generalized eigenvalues (explained in the second item).
The minimum cost for $D=N-1$ (one dimensional reduction) is 
achieved by $N-1$ dominant generalized eigenvectors. This argument
can be justified by noting that the cost of any other subspace
containing $\mathbf{v}_N$ (the generalized eigenvector with the
smallest eigenvalue) can be improved by replacing $\mathbf{v}_N$
with any $\mathbf{v}_n$ which is not already in the span of the
subspace. (This argument, in essence, is the argument utilized to
prove the mean square representation error optimality of the
\textit{Karhunen-Loeve} expansion.) Upon the repeated use of the same
argument, it can be justified that the minimal cost for
$D=\{1,2, \ldots, N\}$ dimensional subspace is achieved by the 
first $D$ dominant generalized eigenvectors.
\item[6.] By using the arguments discussed in previous items, the normalized MSE in (\ref{eqn:nmse_error_cov_orthMPCs}) is minimized when the dominant \textit{generalized eigenvectors} for each MPC are utilized as a dimension reducing beamspace.  
\end{enumerate}

\section*{Appendix II: Calculation of the Error Volume and the Normalized MSE}
\label{sec:Appendix_2} 

\subsection*{Estimation Error Volume}
The error volume, namely, the determinant of the estimation error covariance matrix $\mathbf{R_e}^{mmse}$ in (\ref{eqn:mmse_error_cov}) can be evaluated as  
\begin{align}
&\operatorname{det}\left(\mathbf{R_e}^{mmse}\right)
=\operatorname{det}\left(\mathbf{R}_{full}^{(g)}\right)
\operatorname{det}\left(\mathbf{I}_{K_gL_gN}-
\mathbf{F}_s^{(g)}\left(\mathbf{R}^{(g)}_{\mathbf{y}}\right)^{-1}
\left(\mathbf{F}_s^{(g)}\right)^H\mathbf{R}_{full}^{(g)}\right)
\nonumber \\
&=\operatorname{det}\left(\mathbf{R}_{full}^{(g)}\right)
\operatorname{det}\left(\mathbf{I}_{TD}-
\left(\mathbf{R}^{(g)}_{\mathbf{y}}\right)^{-1}
\left(\mathbf{F}_s^{(g)}\right)^H\mathbf{R}_{full}^{(g)}
\mathbf{F}_s^{(g)}\right)
\nonumber \\
&=\operatorname{det}\left(\mathbf{R}_{full}^{(g)}\right)
\operatorname{det}\left\{\mathbf{I}_{TD}-\left(
\mathbf{I}_{TD}+ 
\left[\left(\boldsymbol{\Upsilon}_S^{(g)}\right)^H
\mathbf{R}^{(g)}_{\boldsymbol{\xi}}
\boldsymbol{\Upsilon}_S^{(g)}\right]^{-1}
\left(\mathbf{F}_s^{(g)}\right)^H\mathbf{R}_{full}^{(g)}\mathbf{F}_s^{(g)}
\right)^{-1} 
\right. \nonumber \\
& \left. \qquad \qquad
\left[\left(\boldsymbol{\Upsilon}_S^{(g)}\right)^H
\mathbf{R}^{(g)}_{\boldsymbol{\xi}}
\boldsymbol{\Upsilon}_S^{(g)}\right]^{-1}
\left(\mathbf{F}_s^{(g)}\right)^H\mathbf{R}_{full}^{(g)}
\mathbf{F}_s^{(g)}\right\} \nonumber \\
&= \operatorname{det}\left(\mathbf{R}_{full}^{(g)}\right)
\operatorname{det}\left\{\mathbf{I}_{TD}- 
\left(\mathbf{I}_{TD}+\sum_{l=0}^{L_g-1}\mathbf{R}^{(g)}_{code}(l) \otimes
\mathbf{SNR}^{(g)}_{mimo}(l)\right)^{-1} 
\left(\sum_{l=0}^{L_g-1}\mathbf{R}^{(g)}_{code}(l) \otimes
\mathbf{SNR}^{(g)}_{mimo}(l)\right)\right\} \nonumber \\
&=\operatorname{det}\left(\mathbf{R}_{full}^{(g)}\right) 
\operatorname{det}\left\{\left(\mathbf{I}_{TD}+\sum_{l=0}^{L_g-1}\mathbf{R}^{(g)}_{code}(l) \otimes \mathbf{SNR}^{(g)}_{mimo}(l)\right)^{-1}
\right\} \nonumber \\
&=\frac{\operatorname{det}\left(\mathbf{R}_{full}^{(g)}\right)}
{\operatorname{det}\left(\mathbf{I}_{TD}+\sum_{l=0}^{L_g-1}\mathbf{R}^{(g)}_{code}(l) \otimes \mathbf{SNR}^{(g)}_{mimo}(l)\right)}.
\label{eqn:error_volume_app}
\end{align}
\noindent In (\ref{eqn:error_volume_app}), the first line follows directly by taking the determinant of the expression given in the last line of (\ref{eqn:mmse_error_cov}). The second line is obtained by using the Sylvester's determinant identity, i.e., 
$\operatorname{det}\left(\mathbf{I} + \mathbf{A}\mathbf{B}\right)=
\operatorname{det}\left(\mathbf{I} + \mathbf{B}\mathbf{A}\right)$.
The third line is obtained after substituting the expressions in (\ref{eqn:matrix_def_error_cov}) into their places, and using the identity 
$\left(\mathbf{A}+\mathbf{B}\right)^{-1}\mathbf{A}=
\left(\mathbf{I} + \mathbf{B}^{-1}\mathbf{A}\right)
\mathbf{B}^{-1}\mathbf{A}$. The fourth line follows from the successive use of the Kronecker product rule by noting the definitions of 
$\mathbf{R}^{(g)}_{code}(l)$ and $\mathbf{SNR}^{(g)}_{mimo}(l)$ matrices in (\ref{eqn:r_code_l}) and (\ref{eqn:snr_mimo_l}). Finally, the fifth line follows from the direct application of the matrix inversion lemma (Woodbury matrix identity), and the sixth line comes from the identity $\operatorname{det}\left(\mathbf{A}^{-1}\right)=
1/\operatorname{det}\left(\mathbf{A}\right)$. 
 
\subsection*{Normalized Error Covariance}
The normalized mean square error can be evaluated as 
\begin{align}
& \operatorname{Tr}\left\{\mathbf{nMSE}^{(g)}\right\} \triangleq \operatorname{Tr}\left\{\mathbb{E}\left\{\left(\mathbf{c}^{(g)}
-\hat{\mathbf{c}}^{(g)}\right)
\left(\mathbf{c}^{(g)}-\hat{\mathbf{c}}^{(g)}\right)^H\right\}\right\}
\nonumber \\
& =\operatorname{Tr}\left\{\mathbf{I}_
{K_g\left(\sum_{l=0}^{L_g-1}r_{g,l}\right)}-
\left(\boldsymbol{\Psi}_D^{(g)}\right)^H\mathbf{W}_{mmse,D}^{(g)}\right\}
\nonumber \\
& =\operatorname{Tr}\left\{\mathbf{I}_{TD}-
\left(\mathbf{R}^{(g)}_{\mathbf{y}}\right)^{-1}
\boldsymbol{\Psi}_D^{(g)}\left(\boldsymbol{\Psi}_D^{(g)}\right)^H\right\}
+\left(K_g\left(\sum_{l=0}^{L_g-1}r_{g,l}\right)-TD\right)
\nonumber \\
& =\operatorname{Tr}\left\{\mathbf{I}_{TD}-
\left(\sum_{l=0}^{L_g-1}\mathbf{R}^{(g)}_{code}(l) \otimes
\mathbf{SNR}^{(g)}_{mimo}(l)+\mathbf{I}_{TD}\right)^{-1}
\left(\sum_{l=0}^{L_g-1}\mathbf{R}^{(g)}_{code}(l) \otimes
\mathbf{SNR}^{(g)}_{mimo}(l)\right)\right\}
\nonumber \\
& \qquad \qquad 
+\left(K_g\left(\sum_{l=0}^{L_g-1}r_{g,l}\right)-TD\right)
\nonumber \\
& =\operatorname{Tr}\left\{
\left(\sum_{l=0}^{L_g-1}\mathbf{R}^{(g)}_{code}(l) \otimes
\mathbf{SNR}^{(g)}_{mimo}(l)+\mathbf{I}_{TD}\right)^{-1}\right\}
+\left(K_g\left(\sum_{l=0}^{L_g-1}r_{g,l}\right)-TD\right).
\label{eqn:nmse_error_cov_app}
\end{align}
\noindent In (\ref{eqn:nmse_error_cov_app}), the first line comes from its definition given in (\ref{eqn:nmse_error_cov_def}), and the second line comes by evaluating the estimation error covariance for 
$\mathbf{c}^{(g)}$ when \textit{Wiener filter} is applied in reduced dimension. The third line comes from the definition of the filter   
$\mathbf{W}_{mmse,D}^{(g)}$ in (\ref{eqn:w_mmse_D}), and the use of the matrix identity $\operatorname{Tr}\left\{\mathbf{A}\mathbf{B}\right\}=
\operatorname{Tr}\left\{\mathbf{B}\mathbf{A}\right\}$. The fourth line follows from successive use of the Kronecker product rule and the definitions of $\mathbf{R}^{(g)}_{code}(l)$ in (\ref{eqn:r_code_l}) and $\mathbf{SNR}^{(g)}_{mimo}(l)$ in (\ref{eqn:snr_mimo_l}) after using the last line of (\ref{eqn:w_mmse_D}), (\ref{eqn:equivalent_channel}) and the following identity 
\begin{equation}
\boldsymbol{\Psi}_D^{(g)}\left[\boldsymbol{\Psi}_D^{(g)}\right]^H=
\left(\mathbf{X}^{(g)} \otimes \left[\mathbf{S}_D^{(g)}\right]^H\right)\left(\mathbf{I}_{K_g} \otimes \mathbf{V}\mathbf{V}^H\right)
\left(\left[\mathbf{X}^{(g)}\right]^H \otimes \mathbf{S}_D^{(g)}\right)
\label{eqn:def_PhiPhi}
\end{equation}
\noindent Finally, the last line of (\ref{eqn:nmse_error_cov_app}) follows from the matrix inversion lemma. 

\section*{Appendix III: High SNR Approximation for RR-MMSE Estimators}
\label{sec:Appendix_3}
\subsection*{Angle-Only Domain}
First, the eigendecomposition of the matrices   
$\mathbf{R}^{(g)}_{code}$ in (\ref{eqn:r_code_total}) and $\mathbf{SNR}^{total,(g)}_{mimo}$ in (\ref{eqn:snr_mimo_total}) can be expressed as
\begin{align}
\mathbf{R}^{(g)}_{code}&=
\sum_{\left\{m \; \vert \beta_m > 0\right\}}
\beta_m \boldsymbol{\phi}_m 
\boldsymbol{\phi}_m^H 
\label{eqn:R_code_svd} \\
\mathbf{SNR}^{total,(g)}_{mimo}&=
\boldsymbol{\Gamma}
\operatorname{diag}\left[\left\{\lambda_n\right\}_{n=1}^{D}\right]
\boldsymbol{\Gamma}^{-1}
\label{eqn:SNR_total_svd}
\end{align}
\noindent where $\boldsymbol{\Gamma} \triangleq \left[\boldsymbol{\gamma}_1 \cdots 
\boldsymbol{\gamma}_D\right]_{D \times D}$. 
If the GEB is adopted as the pre-beamformer, the 
$\mathbf{SNR}^{total,(g)}_{mimo}$ matrix is reduced to a diagonal matrix in the following form (as shown in Appendix I)
\begin{equation}
\mathbf{SNR}^{total,(g)}_{mimo}=
\sum_{\left\{n \; \vert \lambda_n > 0\right\}}\lambda_n
\mathbf{e}_n\mathbf{e}_n^H
\label{eqn:SNR_total_diag}
\end{equation}
\noindent where $\mathbf{e}_n$ is the $D \times 1$ elementary vector where all the entries, except the $n^{th}$ one, are zero.  
By using the eigendecomposition of $\mathbf{R}^{(g)}_{code}$ and 
$\mathbf{SNR}^{total,(g)}_{mimo}$ matrices given in (\ref{eqn:R_code_svd}) and (\ref{eqn:SNR_total_svd}), one can express the \textit{angle} domain RR-MMSE estimator in (\ref{eqn:h_eq_mmse_approx}) as  
\begin{align}
& \hat{\mathbf{h}}_{eff,2}^{(g)}=
\left(\left[\mathbf{X}^{(g)}\right]^H \otimes
\sum_{\left\{n \; \vert \lambda_n > 0\right\}}\lambda_n
\left[\mathbf{e}_n\mathbf{e}_n^H\right]\right) 
\left(\mathbf{I}_{TD}-\sum_{\left\{m,n \; 
\vert \beta_m,\lambda_n > 0\right\}}
\frac{\beta_m\lambda_n}{\beta_m\lambda_n+1}
\left[\boldsymbol{\phi}_m \boldsymbol{\phi}_m^H\right] \otimes
\left[\mathbf{e}_n\mathbf{e}_n^H\right]\right)
\mathbf{y}^{(g)} \nonumber \\
& =\sum_{\left\{n \; \vert \lambda_n > 0\right\}}\lambda_n
\left(\left[\mathbf{X}^{(g)}\right]^H \otimes 
\left[\mathbf{e}_n\mathbf{e}_n^H\right]\right)\mathbf{y}^{(g)}-
\sum_{\left\{m,n \; \vert \beta_m,\lambda_n > 0\right\}}
\frac{\beta_m\left(\lambda_n\right)^2}{\beta_m\lambda_n+1}
\left\{\left(\left[\mathbf{X}^{(g)}\right]^H
\boldsymbol{\phi}_m \boldsymbol{\phi}_m^H\right) \otimes
\left(\mathbf{e}_n\mathbf{e}_n^H\right)\right\}\mathbf{y}^{(g)} 
\nonumber \\
& =\sum_{\left\{m,n \; \vert \beta_m,\lambda_n > 0\right\}}
\frac{\left(\beta_m\right)^{1/2}\lambda_n}{\beta_m\lambda_n+1}
\left(\left[\boldsymbol{\psi}_m \boldsymbol{\phi}_m^H\right] \otimes 
\left[\mathbf{e}_n\mathbf{e}_n^H\right]\right)\mathbf{y}^{(g)}.
\label{eqn:h_eq_mmse_approx_2}
\end{align}
\noindent In (\ref{eqn:h_eq_mmse_approx_2}), the first line follows from obtaining the inverse of the matrix $\mathbf{R}^{(g)}_{code} \otimes \mathbf{SNR}^{total,(g)}_{mimo}$ in (\ref{eqn:h_eq_mmse_approx}). The matrix inverse in (\ref{eqn:h_eq_mmse_approx}) can be expressed in terms of its principal components, in a similar way to the principal components inverse (PCI) technique used in space-time adaptive processing (STAP) in \cite{melvin04}, by noting the fact that $\left\{\boldsymbol{\phi}_m \otimes \boldsymbol{\gamma}_n\right\}_{m,n}$ is the set of eigenvectors for     
$\mathbf{R}^{(g)}_{code} \otimes \mathbf{SNR}^{total,(g)}_{mimo}$ with the corresponding set of positive eigenvalues $\left\{\beta_m\lambda_n\right\}$. The second line follows from the direct multiplication of the terms in two brackets, and the use of the Kronecker product rule by noting that $\mathbf{e}_m^H\mathbf{e}_n=0$ if $m \neq n$. The singular value decomposition (SVD) of $\mathbf{X}^{(g)}$ in (\ref{eqn:complete_training_matrix}) can be written as
\begin{equation}
\mathbf{X}^{(g)}=
\sum_{\left\{m \; \vert \beta_m > 0\right\}}
\left(\beta_m\right)^{1/2} \boldsymbol{\phi}_m 
\boldsymbol{\psi}_m^H 
\label{eqn:X_train_svd}
\end{equation}
\noindent where the $\boldsymbol{\phi}_m$s are the left singular vectors of $\mathbf{X}^{(g)}$ given in (\ref{eqn:R_code_svd}), and    
$\boldsymbol{\psi}_m$s are the right singular vectors. Then, the third line of (\ref{eqn:h_eq_mmse_approx_2}) is obtained after some mathematical manipulations noting that $\boldsymbol{\phi}_m^H\boldsymbol{\phi}_n=0$ for $m \neq n$ by substituting  
$\mathbf{X}^{(g)}$ in (\ref{eqn:X_train_svd}) into its place.  

The \textit{asymptotic high SNR approximation} of (\ref{eqn:h_eq_mmse_approx_2}) can be fulfilled by letting non-zero eigenvalues of the training matrix in (\ref{eqn:R_code_svd}) approach infinity, i.e., $\beta_m \to \infty$ as  
\begin{align}
\hat{\mathbf{h}}_{eff,2}^{(g)} &\approx
\sum_{\left\{m \; \vert \beta_m > 0\right\}}\sum_n
\frac{1}{\left(\beta_m\right)^{1/2}}
\left(\left[\boldsymbol{\psi}_m \boldsymbol{\phi}_m^H\right] \otimes 
\left[\mathbf{e}_n\mathbf{e}_n^H\right]\right)\mathbf{y}^{(g)}
\nonumber \\
&= \left\{\left(\sum_{\left\{m \; \vert \beta_m > 0\right\}}
\frac{1}{\left(\beta_m\right)^{1/2}}
\left[\boldsymbol{\psi}_m \boldsymbol{\phi}_m^H\right]\right)
\otimes \mathbf{I}_D\right\}\mathbf{y}^{(g)} \nonumber \\
&= \left\{ \begin{array}{cc}
\left\{\left(\left[\mathbf{X}^{(g)}\right]^H\mathbf{X}^{(g)}\right)^{-1}\left[\mathbf{X}^{(g)}\right]^H \otimes 
\left[\mathbf{S}_D^{(g)}\right]^H\right\}\mathbf{y} &
\qquad \textrm{ if } T \geq K_gL_g, \\
\\
\left\{\left[\mathbf{X}^{(g)}\right]^H
\left(\mathbf{X}^{(g)}\left[\mathbf{X}^{(g)}\right]^H\right)^{-1} \otimes 
\left[\mathbf{S}_D^{(g)}\right]^H\right\}\mathbf{y} &
\qquad \textrm{ if } T < K_gL_g.
\end{array}\right. 
\label{eqn:h_eq_ml_approx_app}
\end{align}
\noindent In (\ref{eqn:h_eq_ml_approx_app}), the second line follows from the fact that $\sum_n\mathbf{e}_n\mathbf{e}_n^H=\mathbf{I}_D$, and the third line is directly written by recognizing the expression 
$\sum_{\left\{m \; \vert \beta_m > 0\right\}}
\frac{1}{\left(\beta_m\right)^{1/2}}
\left[\boldsymbol{\psi}_m \boldsymbol{\phi}_m^H\right]$ as the \textit{pseudoinverse} of the $\mathbf{X}^{(g)}$ matrix in (\ref{eqn:X_train_svd}) (if it is full-column or row rank), and using (\ref{eqn:space_time_noise_reduced}).

\subsection*{Joint Angle-Delay Domain}
By using the eigendecomposition of $\mathbf{R}^{(g)}_{code}(l)$ and 
$\mathbf{SNR}^{(g)}_{mimo}(l)$ in (\ref{eqn:R_code_l_svd}) and (\ref{eqn:SNR_l_svd}), the joint \textit{angle-delay} domain RR-MMSE estimator in (\ref{eqn:h_eq_mmse_est}) can be expressed in an explicit form. First, the GEB is adopted as the dimension reducing pre-beamformer. In this case, the $\mathbf{SNR}^{(g)}_{mimo}(l)$ matrix is reduced to the following diagonal matrix approximately 
(as shown in Appendix I): $\mathbf{SNR}^{(g)}_{mimo}(l)=\lambda^l\mathbf{E}_{D,l}$ when the rank of each MPC covariance is one and $D=L_g$. Then, the following rank-1 approximation of the estimator is obtained: 
\begin{align}
& \hat{\mathbf{h}}_{eff}^{(g)} \approx
\left(\sum_{l=0}^{L_g-1} \mathbf{X}^{(g)} \left[\mathbf{I}_{K_g} \otimes \mathbf{E}_{L_g,l}\right]
\otimes \mathbf{SNR}^{(g)}_{mimo}(l) \right)^H 
\nonumber \\
& \qquad
\left(\sum_{l=0}^{L_g-1}
\sum_{\left\{m \; \vert \beta_m^l > 0\right\}}\beta_m^l
\boldsymbol{\phi}_m^l \left[\boldsymbol{\phi}_m^l\right]^H
\otimes \lambda^l\mathbf{E}_{D,l} + \mathbf{I}_{TD}\right)^{-1}
\mathbf{y}^{(g)} \nonumber \\
& =\left(\sum_{l=0}^{L_g-1} \left(\mathbf{X}^{(g)} \left[\mathbf{I}_{K_g}
\otimes \mathbf{E}_{L_g,l}\right]\right)^H 
\otimes \lambda^l\mathbf{E}_{D,l}\right)
\left( \mathbf{I}_{TD}- 
\sum_{l=0}^{L_g-1}
\sum_{\left\{m \; \vert \beta_m^l > 0\right\}}
\frac{\beta_m^l\lambda^l}{\beta_m^l\lambda^l+1}
\boldsymbol{\phi}_m^l \left[\boldsymbol{\phi}_m^l\right]^H
\otimes \mathbf{E}_{D,l}\right)\mathbf{y}^{(g)}\nonumber \\
& =\sum_{l=0}^{L_g-1} \lambda_l\left\{
\left(\mathbf{X}^{(g)} \left[\mathbf{I}_{K_g}
\otimes \mathbf{E}_{L_g,l}\right]\right)^H \otimes \mathbf{E}_{D,l}\right\}
\mathbf{y}^{(g)} 
\nonumber \\
& \qquad
-\sum_{l=0}^{L_g-1}\sum_{\left\{m \; \vert \beta_m^l > 0\right\}}
\frac{\beta_m^l\left(\lambda^l\right)^2}{\beta_m^l\lambda^l+1} 
\left\{\left(\mathbf{X}^{(g)} \left[\mathbf{I}_{K_g}
\otimes \mathbf{E}_{L_g,l}\right]\right)^H
\boldsymbol{\phi}_m^l \left[\boldsymbol{\phi}_m^l\right]^H
\otimes \mathbf{E}_{D,l}\right\}\mathbf{y}^{(g)}\nonumber \\
& =\sum_{l=0}^{L_g-1}\sum_{\left\{m \; \vert \beta_m^l > 0\right\}}
\frac{\left(\beta_m^l\right)^{1/2}\lambda^l}{\beta_m^l\lambda^l+1}
\left(\boldsymbol{\psi}_m^l \left[\boldsymbol{\phi}_m^l\right]^H
\otimes \mathbf{E}_{D,l}\right)\mathbf{y}^{(g)} 
\label{eqn:h_eq_mmse_approx_orthMPCs}
\end{align}
\noindent In (\ref{eqn:h_eq_mmse_approx_orthMPCs}), the second line follows from the fact that for different $l$, $\mathbf{R}^{(g)}_{code}(l) \otimes
\mathbf{SNR}^{(g)}_{mimo}(l)$ have orthogonal eigenspaces as explained in Section \ref{sec:nearly_opt_solution}. In this case, it can be shown that $\bigoplus_{l=0}^{L_g-1}\left\{\boldsymbol{\phi}_m^l \otimes \mathbf{e}_l\right\}$ (orthogonal direct sum) forms the eigenvectors of 
$\sum_{l=0}^{L_g-1}\mathbf{R}^{(g)}_{code}(l) \otimes
\mathbf{SNR}^{(g)}_{mimo}(l)$ with the corresponding set of positive eigenvalues $\left\{\beta_m^l\lambda^l\right\}$ when $\mathbf{e}_l\mathbf{e}_l^H=\mathbf{E}_{D,l}$. Then, by using the PCI technique, the inverse of the matrix inside the second bracket can be evaluated.   
The third line follows from direct multiplication of the terms in two brackets, and the use of the Kronecker product rule by noting that $\mathbf{E}_{D,l_1}^H\mathbf{E}_{D,l_2}=0$ if $l_1 \neq l_2$. The singular value decomposition (SVD) of $\left(\mathbf{X}^{(g)} \left[\mathbf{I}_{K_g} \otimes \mathbf{E}_{L_g,l}\right]\right)$ in
(\ref{eqn:h_eq_mmse_approx_orthMPCs}) can be written as
\begin{equation}
\mathbf{X}^{(g)} \left[\mathbf{I}_{K_g} \otimes \mathbf{E}_{L_g,l}\right]=
\sum_{\left\{m \; \vert \beta_m^l > 0\right\}}
\left(\beta_m^l\right)^{1/2} 
\boldsymbol{\psi}_m^l \left[\boldsymbol{\phi}_m^l\right]^H
\label{eqn:X_train_l_svd}
\end{equation}
\noindent Then, the fourth line of (\ref{eqn:h_eq_mmse_approx_orthMPCs}) is obtained after some mathematical manipulations noting that 
$\left(\boldsymbol{\phi}_m^l\right)^H\boldsymbol{\phi}_n^l=0$ for $m \neq n$ by substituting (\ref{eqn:X_train_l_svd}) into its place.  

The following \textit{asymptotic high SNR approximation} of (\ref{eqn:h_eq_mmse_approx_orthMPCs}) can be obtained by letting non-zero eigenvalues of the $\mathbf{R}^{(g)}_{code}(l)$ in (\ref{eqn:R_code_l_svd}) approach infinity, i.e., $\beta_m^l \to \infty$ for all $m$ and $l=0,\ldots,L_g-1$   
\begin{align}
\hat{\mathbf{h}}_{eff}^{(g)} &\approx
\left\{\sum_{l=0}^{L_g-1}\left(\sum_{\left\{m \; \vert \beta_m^l > 0\right\}}
\frac{1}{\left(\beta_m^l\right)^{1/2}}
\boldsymbol{\psi}_m^l \left[\boldsymbol{\phi}_m^l\right]^H\right)
\otimes \mathbf{E}_{D,l}\right\}\mathbf{y}^{(g)} \nonumber \\
& =\sum_{l=0}^{L_g-1}
\left( \operatorname{pinv}\left\{\mathbf{X}^{(g)} 
\left[\mathbf{I}_{K_g} \otimes \mathbf{E}_{L_g,l}\right]\right\}
\otimes \left[\mathbf{S}_D^{(g)}\mathbf{E}_{D,l}\right]^H\right)\mathbf{y}. 
\label{eqn:h_eq_ml_approx_orthMPCs_app}
\end{align}
\noindent In (\ref{eqn:h_eq_ml_approx_orthMPCs_app}), the second line follows by recognizing the expression 
$\sum_{\left\{m \; \vert \beta_m^l > 0\right\}}
\frac{1}{\left(\beta_m^l\right)^{1/2}}
\boldsymbol{\psi}_m^l \left[\boldsymbol{\phi}_m^l\right]^H$ as the \textit{pseudoinverse} of $\mathbf{X}^{(g)} \left[\mathbf{I}_{K_g} \otimes \mathbf{E}_{L_g,l}\right]$ in (\ref{eqn:X_train_l_svd}) (when the matrix is rank deficient), and using   
(\ref{eqn:space_time_noise_reduced}).

\bibliographystyle{ieeetr}
\bibliography{SC_MIMO_MMW}
\end{document}